\documentclass[intlimits,sumlimits,12pt]{iopart}

\expandafter\let\csname equation*\endcsname\relax
\expandafter\let\csname endequation*\endcsname\relax

\newcommand{\inte}{\int\limits}

\newcommand{\summ}{\sum\limits}

\newcommand{\diag}{\textrm{diag\,}}

\def\newblock{\hskip .11em plus.33em minus.07em}

\usepackage{iopams}
\usepackage{amsmath}
\usepackage{dsfont}
\usepackage{cite}
\usepackage{graphicx}
\usepackage[numbers]{natbib}

\begin{document}

\title[Exact Multivariate Amplitude Distributions]{Exact Multivariate
  Amplitude Distributions for Non--Stationary Gaussian or Algebraic
  Fluctuations of Covariances or Correlations}

\author{Thomas Guhr and Andreas Schell}

\address{Fakult\"at f\"ur Physik, Universit\"at Duisburg--Essen, Duisburg, Germany}
\ead{thomas.guhr@uni-due.de}
\vspace{10pt}

\begin{abstract}
  Complex systems are often non--stationary, typical indicators are
  continuously changing statistical properties of time series. In
  particular, the correlations between different time series
  fluctuate.  Models that describe the multivariate amplitude
  distributions of such systems are of considerable
  interest. Extending previous work, we view a set of measured,
  non--stationary correlation matrices as an ensemble for which we set
  up a random matrix model. We use this ensemble to average the
  stationary multivariate amplitude distributions measured on short
  time scales and thus obtain for large time scales multivariate
  amplitude distributions which feature heavy tails. We explicitly
  work out four cases, combining Gaussian and algebraic
  distributions. The results are either of closed forms or single
  integrals. We thus provide, first, explicit multivariate
  distributions for such non--stationary systems and, second, a tool
  that quantitatively captures the degree of non--stationarity in the
  correlations.
\end{abstract}

%
\vspace{2pc}
\noindent{\it Keywords}: non--stationarity, multivariate statistics, algebraic heavy tails, random matrix theory

\submitto{\JPA}

%
%

\section{Introduction}
\label{sec0}

A wealth of complex systems show non--stationarity as characteristic
feature, \textit{i.e.} they lack any kind of
equlibrium~\cite{Gao1999,Hegger2000,Bernaola2001,Rieke2002}. Non--stationarity
shows up in many different ways~\cite{Zia2004,Zia2006}, for example,
the increasing share of wind energy fed into the power grid is greatly
appreciated from an environmental viewpoint, but causes problems for
the stability of the grid due to non--stationarity and the lack of
predictability for wind speed and wind direction. In
electroencephalography (EEG) electrical currents are recorded at
different positions on the scalp to measure the brain activity. The
correlations strongly depend on the overall state of the
brain~\cite{pijn1991chaos,Mueller2005}. Wave packets traveling through
disordered systems also show non--stationarities, even for static
disorder.  The time series of wave intensities at different positions
change with the direction or the composition of the wave packet which
then modifies the
correlations~\cite{Hohmann2010,metzger2014statistics,degueldre2015random}. Finance
also features this type of non--stationarity because the business
relations between the firms and the traders' market expectations
change, the non--stationarity becomes dramatic in the state of
crisis~\cite{bekaert1995,Longin1995,Onnela2003,Zhang2011,Song2011,Sandoval2012,Munnix2012}.
Further examples are found in many complex systems, such as velocity
fluctuations in turbulent flows, heartbeat dynamics, series of waiting
times, etc. \cite{Ghasemi2006, PhysRevE.87.062139, PhysRevE.75.060102,
  Schafer20103856}.  Although most approaches of statistical physics
based on the existence of equilibrium, stationarity or
quasi--stationarity are not applicable, these systems pose questions
similar to the ones in equlibrium. Is there generic, universal
behavior and how can we identify it? --- What are proper statistical
models, \textit{i.e.}  models based on the assumption of randomness
for some parts or aspects of the systems? --- Sometimes other
approaches, \textit{e.g.} from many--body physics carry over or
provide useful inspiration. This is especially so for the interplay
between coherent, collective motion and incoherent motion of the
individual
particles~\cite{bohr1969nuclear,zelevinsky1996quantum,Guhr1998}.

Here, we have two goals. First, we construct analytical results for
the multivariate distributions of amplitudes, measured as time series
in correlated, non--stationary systems. Second, we thereby provide
quantitative measures for the degree of non--stationarity in the
correlations. We use the word ``amplitude'' to refer to any observable,
directly measured or inferred from a measurement, at a given position
and a given time. The word ``position'' is meant in a general sense, a
geographical point, a location, a specific stock and so on. The
amplitudes at all times of measurement at a given position form the
corresponding time series. The challenge we master is to capture the
non--stationarity of the correlations between these time series in
terms of a statistical ensemble. We considerably extend a random
matrix approach to tackle these issues that we presented some years
ago~\cite{Schmitt2013}, see also some modification in
Ref.~\cite{Meudt2015}. It also turned out useful in the context of
credit risk and its impact on systemic
stability~\cite{schmitt2014credit,schmitt2015credit}. Relations to
other systems were discussed in Ref.~\cite{SBGSK2015}. In
Ref.~\cite{Schmitt2013}, we showed that the fluctuations of the
correlations lift the tails of the multivariate amplitude
distributions, making them heavy--tailed. Our new results extend that.

Our model may be viewed as a justification of compounding or mixture
techniques in mathematics and superstatistics in
physics~\cite{dubey1970compound,10.2307/1402598,Beck2003,Abul-Magd2009,Doulgeris2010,Forbes2014,Abul-Magd2009}.
By tracing heavy--tailed distributions back to fluctuations of the
correlations, we explain such ad--hoc approaches.  However, our
previous studies were based on a Gaussian assumption for the
statistical ensemble. To avoid misunderstandings, we underline that
this Gaussian assumption does not at all imply Gaussian form of the
multivariate amplitude distribution, rather, its tails are
exponential~\cite{Schmitt2013}. Here, we extend this by also
considering ensembles of algebraic covariance or correlation matrices
in the form of certain determinants.  This is relevant, as many
complex systems, not only financial markets~\cite{bouchaud2000theory},
are known to have algebraically heavy--tailed distributions. Our
earlier results~\cite{Schmitt2013} show a discrepancy from the
exponential behavior way out in the tails between data and model which
is due to the above mentioned Gaussian assumption. However, even
though the data analysis triggered the present study of algebraic
covariance or correlation matrix distributions, the detailed
discussion of the empirical aspects addresses a different community
and will thus be published elsewhere.

A quite remarkable feature of the multivariate distributions that we
derive here is that, eventually, they are of closed form or involve
only single integrals. Furthermore, the number of free parameters is
low: there is one parameter, measuring how strongly the
non--stationary correlations fluctuate, and one or two shape
parameters for the tails. We provide a technique to determine
them. All other parameters can be directly measured from the data.
These features were major motivations for our model, as many
descriptions of statistical observables in complex systems are based on
ad--hoc fit formulas with parameters that lack a clear interpretation
from a data viewpoint.

Random matrix models~\cite{Mehta2004,Guhr1998} fall, in the context of
the present discussion, into two classes: (I) The ensemble is
fictitious. It comes into play via an ergodicity argument only.  (II)
The ensemble really exists and can be identified in the system. The
issue of ergodicity does not arise. --- The vast majority of random
matrix models lies in class (I). It is conceptually important that we
here present an random matrix model in class (II) which may be seen as
a new interpretation of the Wishart model and generalizations thereof
for random covariance or correlation matrices~\cite{Wishart1928}. As
financial markets belong to the complex systems triggering our earlier
work~\cite{Schmitt2013}, it is worth mentioning the numerous random
matrix applications in
finance~\cite{bouchaud2000theory,Laloux1999,Laloux2000,Plerou1999a,Plerou2002,Pafka2004,potters2005financial,drozdz2008empirics,Kwapien2006,biroli2007student,Burda2001,Burda2002,Akemann2008,burda2011applying},
including non--Gaussian ensembles. To the best of our knowledge, all
of them fall into class (I) and focus on other observables, while we
here take a different route. Eventually, we arrive at rather universal
and generic results for our distributions, supporting our view that
non--stationarities can lead to universal features.

The paper is organized as follows. In Sec.~\ref{sec1}, we pose the
problem and develop the startegy of our model. We do that in some
detail and, hopefully, in sufficient clarity, because our earlier
work~\cite{Schmitt2013} seems to have been misunderstood in parts of
the community. We devote Sec.~\ref{sec1a} to the mathematical methods
which allows us to perform the ensuing calculations in a compact,
modular style. In Sec.~\ref{sec2}, we derive the results for the
multivariate ensemble averaged amplitude distributions. To facilitate
the analysis of data in forthcoming studies, we develop in
Sec.~\ref{sec3} a technique to compare these multivariate
distributions efficiently with data. We present our conclusions in
Sec.~\ref{sec4}. Some details are relegated to the Appendix.

\section{Posing the Problem and Developing the Strategy}
\label{sec1}

To set the tone and to introduce our notation, we briefly review
basics on covariances and correlations in Sec.~\ref{sec11}. After a
discussion of non--stationarity in Sec.~\ref{sec12}, we present the
key ideas of our random matrix model in Sec.~\ref{sec13} and choose in
Sec.~\ref{sec14} explicit forms of the distributions which are the
ingredients of our model.

\subsection{Covariances and Correlations for Time and Position Series}
\label{sec11}

Suppose we measured amplitudes $A_k(t)$ at $K$ positions labeled
$k=1,\ldots,K$ and at equidistant times $t=1,\ldots,T$. These
amplitudes can be any type of data, temperatures, water levels,
electrical voltages, stock prices and so on. The term ``position'' is
used in a very general way, it can be a geographical point, when
\textit{e.g.}  water levels of rivers are measured or a location on
the scalp of a person in EEG measurements, or in an abstract sense a
company when stock prices are considered. We always assume that time
is normalized such that $t$ is an integer and $T$ is the total number
of points in time.  Keeping with a common notation, we write the
postition as an index and the time as the argument of the data
$A_k(t)$, other notations such as $A(k,t)$ or $A_{kt}$ would be
equivalent. Our data may be ordered in the rectangular $K\times T$
matrix
\begin{align}
	A&=\begin{bmatrix}
			A_1 (1) & A_1 (2) & \cdots & A_1 (T) \\
			A_2 (1) & A_2 (2) & \cdots & A_2 (T) \\
			\vdots & \vdots & \ddots & \vdots \\
			A_K (1) & A_K (2)  & \cdots & A_K (T) \\
	\end{bmatrix} \ .
        \label{eq:daten}
\end{align}
The rows of this matrix are the well known time series $A_k(t),
t=1,\ldots,T$ which contain the amplitudes at the same position for
all times. Importantly, the columns may be interpreted in a similar,
but dual, way: they are the position series which collect the
amplitudes at the same time for all positions. These two types of
series provide different information, particularly on covariances and
correlations. To obtain the covariances of the time series, we first
have to substract their mean values in time,
\begin{align}
\langle A_k \rangle_T = \frac{1}{T} \sum_{t=1}^T A_k(t) \ ,
\label{mvt}
\end{align}
which yields the normalized time series
\begin{align}
B_k(t) = A_k(t) - \langle A_k \rangle_T \ , \qquad t=1,\ldots,T 
\label{nt}
\end{align}
that are the rows of the $K\times T$ matrix $B$. The covariances of the time series are then
the normalized scalar products
\begin{align}
  \Sigma_{kl} = \langle B_k B_l \rangle_T = \frac{1}{T} \sum_{t=1}^T B_k(t)B_l(t) \ .
\label{covt}
\end{align}
In the statistics literature, one often uses the prefactor $1/(T-1)$
and refers to the covariances \eqref{covt} as biased. In physics, the
present choice seems more common.  The $\Sigma_{kl}$ are the elements
of the $K\times K$ sample covariance matrix of the time series
\begin{align}
  \Sigma = \frac{1}{T} BB^\dagger \ .
\label{covtm}
\end{align}
We always use the dagger symbol to indicate the matrix transpose. We define
the variances of the time series
\begin{align}
  \sigma_{k} = \Sigma_{kk} \qquad \textrm{with} \qquad \sigma = \diag(\sigma_1,\ldots,\sigma_K)
\label{vart}
\end{align}
being a diagonal matrix. All variance are positive definite and
\begin{align}
  C = \sigma^{-1} \Sigma \sigma^{-1} 
\label{corrtm}
\end{align}
is the correlation matrix of the time series.

However, when studying the covariances of the position series, we must
apply another normalization employing the mean values of the columns,
\begin{align}
\langle A(t) \rangle_K = \frac{1}{K} \sum_{k=1}^K A_k(t) \ ,
\label{mvk}
\end{align}
leading to the normalized position series
\begin{align}
\widetilde{B}_k(t) = A_k(t) - \langle A(t) \rangle_K \ , \qquad k=1,\ldots,K \ . 
\label{nk}
\end{align}
These are the colums of the $K\times T$ matrix $\widetilde{B}$. The covariances of
the position series are the normalized scalar products
\begin{align}
\Xi(t,s) = \langle \widetilde{B}(t)\widetilde{B}(s) \rangle_K = \frac{1}{K} \sum_{k=1}^K \widetilde{B}_k(t)\widetilde{B}_k(s) 
\label{covk}
\end{align}
which form the $T\times T$ covariance matrix of the position series
\begin{align}
  \Xi = \frac{1}{K} \widetilde{B}^\dagger \widetilde{B} \ .
\label{covkm}
\end{align}
Introducing the variances of the position series
\begin{align}
  \xi(t) = \Xi(t,t) \ , \quad \textrm{and} \qquad \xi = \diag(\xi(1),\ldots,\xi(T)) \ ,
\label{vark}
\end{align}
we may define
\begin{align}
  D = \xi^{-1} \Xi \xi^{-1} 
\label{corrkm}
\end{align}
as the correlation matrix of the position series.  The covariances
$\Sigma_{kl}$ quantify the differences or similarity of two different
time series at equal times, while the covariances $\Xi(t,s)$ give
information on what happens at the same positions for different times
and thus on possible non--Markovian behavior of the system under
consideration. By their very definition, $\Sigma$ and $\Xi$ and thus
$C$ and $D$ are positive semi--definite, they are positive definite
for $K\le T$ or $T\le K$, respectively.  For uncorrelated time series,
one has $C=\mathds{1}_K$, \textit{i.e.}  the $K\times K$ unit matrix, Markovian
data are characterized by $D=\mathds{1}_T$. It is worth mentioning that the
covariance matrices $\Sigma$ and $\Xi$ have physical dimensions,
namely the square of the dimension carried by the amplitudes, while the
correlation matrices $C$ and $D$ are dimensionless. As this will
become relevant in the later discussion, we continue to work for the
time being with both, the covariances and the correlations.

In the statistics literature, the sample covariance and correlation
matrices are viewed as estimators of the population covariance or
correlation matrices. The idea behind is that every sample is a subset
of the full data set, \textit{e.g.} the income distribution for a
whole country is not obtained by asking everyone who receives a
salary, it is estimated from a sample, \textit{i.e.}  from a subset of
the population that is considered sufficiently large. This issue,
however, is only of limited relevance in the present context. We have
systems in mind for which all time series are at our disposal,
\textit{i.e.} the sample matrices yield very reliable
estimators. Rather, it is the issue of non--stationarity that we will
tackle here.

\subsection{Non--Stationarity}
\label{sec12}

\begin{figure}[htbp]
  \begin{center}
    \includegraphics[width=0.32\textwidth]{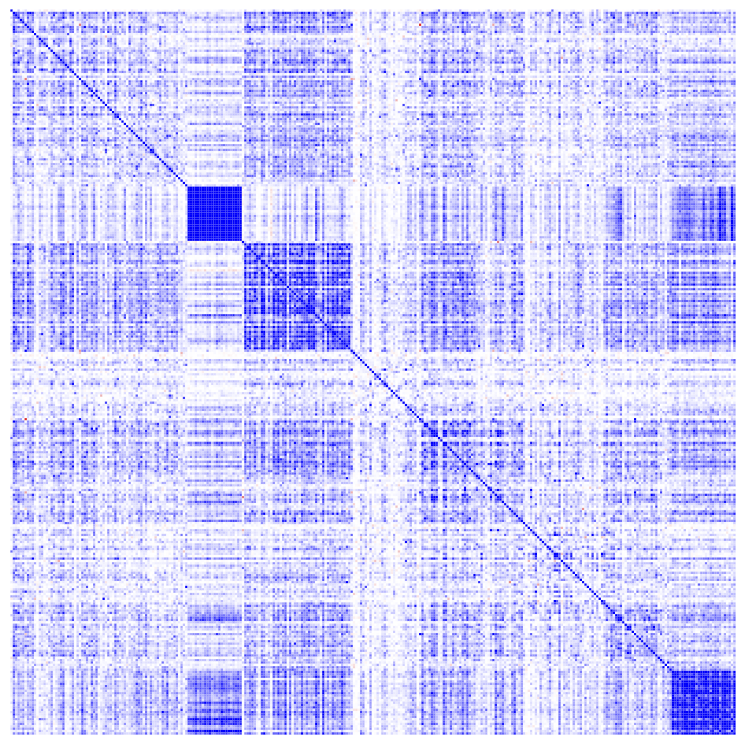}
    \includegraphics[width=0.32\textwidth]{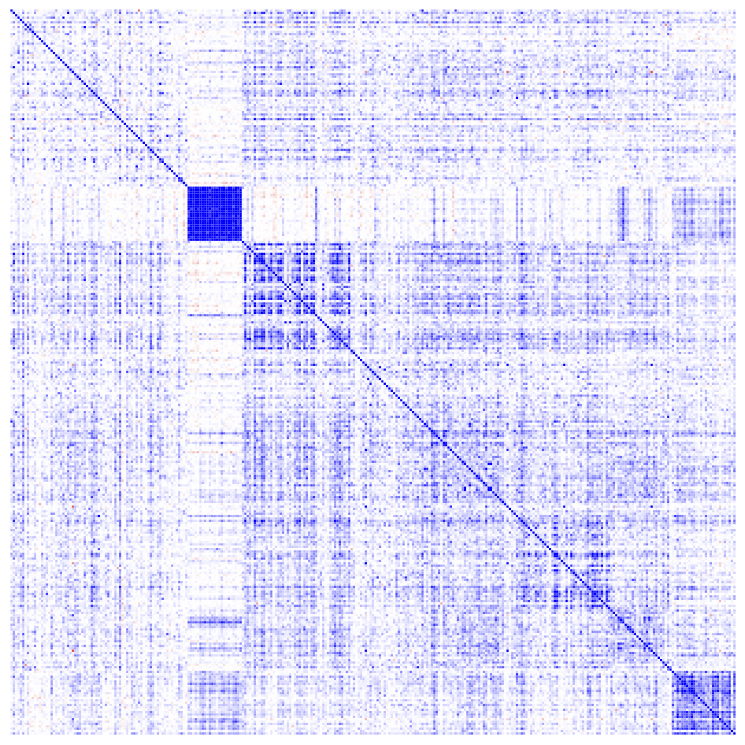}
  \end{center}
  \caption{Correlation matrices of $K=306$ companies in the Standard
    and Poor's 500 index for the fourth quarter of 2005 and the first
    quarter of 2006, the darker, the stronger the correlation. The
    companies are sorted according to industrial sectors. Taken from
    Ref.~\cite{Schmitt2013}.}
 \label{fig1}
\end{figure}

\begin{figure}[htbp]
  \begin{center}
    \phantom{abc}
    \vspace{0.6cm}
    \includegraphics[width=0.7\textwidth]{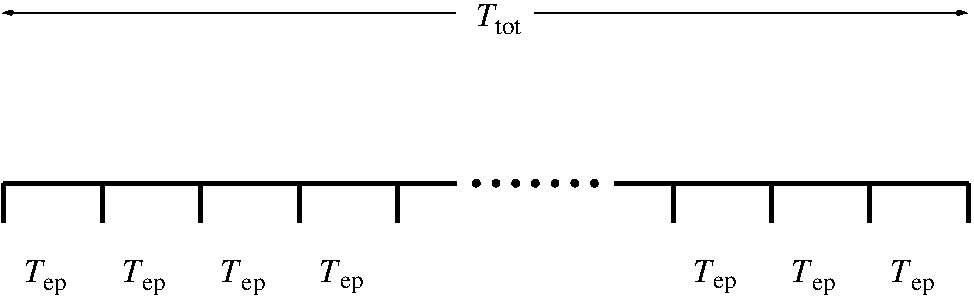}
  \end{center}
  \caption{Large time interval of length $T_\textrm{tot}$, divided
    into $N_\textrm{ep}$ epochs of length $T_\textrm{ep}$.}
 \label{fig2}
\end{figure}

In a non--stationary complex system, crucial parameters or
distributions of observables change in an erratic, unpredictable way
over time. Among the examples mentioned in Sec.~\ref{sec0},
correlations in financial markets are particulary illustrative.
Figure~\ref{fig1} shows two large correlation matrices of
stock prices for companies in the Standard and Poor's 500 index
ordered according to the industrial sectors. The time series were
measured in subsequent quarters. Two important features are
visible. First, the matrices look clearly different, because the
business relations between the companies and the market expectations
of the traders change in time. Second, the coarse structure remains
similar, indicating some stability of the industrial
sectors. Subsequently measured correlation matrices for other complex
systems would have similar properties. The question is how the
non--stationarity, \textit{i.e.} the fluctuations of the correlations
on shorter time scales influence the statistics of a complex system
studied over long time scales. Specifically, we aim at developing a
model for the multivariate distribution of amplitudes $r_k,
\ k=1,\ldots,K$ ordered in the $K$ component vector
$r=(r_1,\ldots,r_K)$. In our notation, we distinguish the amplitudes
$A_k(t)$, measured at a particular time $t$ from the amplitudes $r_k$,
sampled over some time interval. As in our previous
study~\cite{Schmitt2013}, we assume that we may divide the data
measured over a large time scale $T_\textrm{tot}$ into epochs of
length $T_\textrm{ep}$ within which the system is to a good
approximation stationary, while it changes from epoch to epoch as in
the example of the financial market in Fig.~\ref{fig1}. This time
scale should be chosen in such a way that the number
\begin{align}
N_\textrm{ep} = \frac{T_\textrm{tot}}{T_\textrm{ep}}
\label{nep}
\end{align}
of epochs is integer.  While a proper determination of the time scale
$T_\textrm{ep}$ is essential for data analysis, we emphasize that this
time scale does not directly enter our model to be set up in the
sequel. Figure~\ref{fig2} illustrates our construction which
interprets the non--stationary system studied over the long time
$T_\textrm{tot}$ as assembled of $N_\textrm{ep}$ subsequent systems
which are all approximately stationary on their time scale
$T_\textrm{ep}$.

\subsection{Random Matrix Model for a Truly Existing Data Ensemble}
\label{sec13}

Random Matrix Theory usually relies on a concept which is sometimes
referred to as second ergodicity, these are the models in class (I),
as defined in Sec.~\ref{sec0}. Ergodicity is an indispensable feature
of statistical mechanics, as it states the equivalence of time and
ensemble averages. It is worth underlining that the ensembles in
statistical mechanics are fictitious, a powerful mathematical
construction to facilitate, by means of ergodicity, the physically
relevant time average. When using random matrices to model spectral
statistics of individual systems such as a chaotic billiard or the
celebrated Hydrogen atom in a strong magnetic field, a very similar
line of reasoning is applied. Empirically, the spectral statistics of
the individual system is obtained by spectral averages of one single
spectrum which, to allow for a meaningful statement, has to contain a
large number of levels or resonances. Second ergodicity, first proven
in Ref.~\cite{French1978}, states that such a spectral average is
equivalent to an average over an ensemble of random matrices, provided
their dimension is very large. Again, this ensemble is fictitious as
one is interested in comparing with the spectral statistics of one
individual system. The setup and the previous applications of the
Wishart model for random covariance or correlation
matrices~\cite{Wishart1928} were also in class (I).  However, there
are some exceptions, in which the spectral statistics of a truly
existing ensemble of systems described by a Hamilton or Dirac operator
is compared with the one of a random matrix ensemble that is not
fictitious, leading to models in class (II). Important examples are
the random matrix analysis of the nuclear data ensemble~\cite{Haq1982}
which combines data measured for a whole set of nuclei as well as
chiral random matrix theory~\cite{Jac00} modeling lattice gauge
calculations which evaluate spectra for quarks propagating on the
lattice for a whole set of different gauge field configurations. We
here proceed in a similar spirit for random matrix models of
covariances or correlations, extending our earlier
studies~\cite{Schmitt2013}.

The non--stationarity that we wish to model is fully encoded in the
set of the $N_\textrm{ep}$ different matrices $\Sigma_\textrm{ep}$ or
$C_\textrm{ep}$. We may view either set of $N_\textrm{ep}$ matrices as
a matrix ensemble, and our goal is to model it by an ensemble of
random matrices. This observation will lead us to a random matrix
model in class (II). We model the truly existing ensemble of positive
definite matrices $\Sigma_\textrm{ep}$ or $C_\textrm{ep}$ by random
ones,
\begin{align}
  \Sigma_\textrm{ep} \longrightarrow \frac{1}{N} XX^\dagger
  \qquad \textrm{or} \qquad
C_\textrm{ep} \longrightarrow \frac{1}{N} XX^\dagger \ ,
\label{reprmt}
\end{align}
where the form of the right hand sides ensures positive
semi--definiteness. The random matrices $X$, the model data matrices,
have dimension $K \times N$ where the number of rows $K$ is determined
by the fact that the matrices $\Sigma_\textrm{ep}$ and $C_\textrm{ep}$
have dimension $K \times K$. The number of columns $N$, however, is
not fixed by the relation \eqref{reprmt}, it may be interpreted as
length of the random model time series which form the rows of $X$. It
is a free parameter in our construction and is, at the end of our
calculations, not even restricted to integer values. To provide
already now an intuitive and plausibel interpretation and anticipating
the later discussion, we mention that the amount of information on a
system grows with $T$, the longer the time series, the lower are the
fluctuations when measuring averages. Accordingly, $N$ in our model
characterizes the strength of the fluctuations in the ensemble of the
matrices $\Sigma_\textrm{ep}$ or $C_\textrm{ep}$ around the matrix
mean values $\Sigma_\textrm{tot}$ or $C_\textrm{tot}$, respectively,
for the long time interval. We notice that the truly measured matrices
are strongly correlated, the stronger, the closer in time their epochs
are. Our model is not meant to capture those correlations directly,
rather, it is set up to model the fluctuations of the matrices
$\Sigma_\textrm{ep}$ or $C_\textrm{ep}$ in different epochs around
their mean values, which implicitly takes care of the mentioned
correlations across the epochs. This is fully consistent with the
basic idea of statistical mechanics and in the present context best
understood when comparing with chiral random matrices.

Motivated by our data analysis to be published elsewhere, we assume
that the analytical forms of the multivariate amplitude distributions
$p(r|\Sigma_\textrm{ep})$ or $p(r|C_\textrm{ep})$ for a given epoch
have the same functional form for all epochs.  Nevertheless, they
differ from epoch to epoch. We assume that this variation may be fully
captured by the positive definite matrices $\Sigma_\textrm{ep}$, which
have the dimension of the amplitudes squared, or $C_\textrm{ep}$,
which are dimensionless, respectively, and differ from epoch to epoch.
Importantly, these matrices can only be estimated directly by sample
covariance or correlation matrices if $p(r|\Sigma_\textrm{ep})$ or
$p(r|C_\textrm{ep})$ have Gaussian form. For other functional forms,
$\Sigma_\textrm{ep}$ or $C_\textrm{ep}$ cannot directly be obtained in
this way, but it is essential to construct our model in such a way
that they are related to the sample covariance or correlation matrices
and it must be possible to determine them from those.

In the sequel, we focus on the correlation matrices, but later on we
will show how everything easily carries over to covariance
matrices. We draw the random data matrices $X$ and thus the model
matrices $XX^\dagger/N$ from a distribution $w(X|C,D)$. It
parametrically depends on a positive definite matrix $C$ which,
similary to the above discussion, only coincides with the $K \times K$
sample correlation matrix of time series measured over the long time
$T_\textrm{tot}$ for a Gaussian choice of $w(X|C,D)$, but has to be
related to the sample correlation matrix in other cases. It is an
intrinsic property of our model that the second input matrix, the
matrix $D$, modeling the correlations of the position series, has
dimension $N\times N$. The crucial concept now consists of carrying
out the random matrix ensemble average
\begin{align}
\langle p\rangle(r|C,D) = \int p\left(r\Big|\frac{1}{N} XX^\dagger\right) w(X|C,D) \textrm{d}[X] \ ,
\label{rmteac}
\end{align}
in which the replacement \eqref{reprmt} is used in the multivariate
distributions $p(r|C_\textrm{ep})$. The
invariant measure or volume element
\begin{align}
\textrm{d}[X] = \prod_{k=1}^K \prod_{n=1}^N \textrm{d}X_k(n)
\label{rmtvol}
\end{align}
is simply the product of the differentials of all independent matrix
elements.  The ensemble average \eqref{rmteac} yields the distribution
for all amplitude data measured over the long time $T_\textrm{tot}$
subjected to the non--stationarity of the covariances.

\subsection{Choice of Amplitude and Ensemble Distributions}
\label{sec14}

Motivated by data analyses, we choose two forms of the multivariate
amplitude distribution in each of the epochs. As in
Ref.~\cite{Schmitt2013}, we employ the multivariate Gaussian
\begin{align}
  p_G(r|C_\textrm{ep}) &= \frac{1}{\sqrt{\det 2\pi C_\textrm{ep}}}
                       \exp\left(-\frac{1}{2} r^\dagger C^{-1}_\textrm{ep} r\right)
\label{eq:GauVer}
\end{align}
and, as a new choice to model heavy tails, the algebraic distribution
\begin{align}
  p_A(r|C_\textrm{ep}) &= \frac{\alpha_{K1lm}}{\displaystyle\left(1+\frac{1}{m} r^\dagger C^{-1}_\textrm{ep} r\right)^l} \nonumber\\
  \alpha_{K1lm} &= \sqrt{\frac{2}{m}}^K \frac{\Gamma(l)}{\Gamma(l-K/2)}
                                              \frac{1}{\sqrt{\det 2\pi C_\textrm{ep}}}                           
\label{eq:DetVer}
\end{align}
with a shape determined by the parameters $l$ and $m$. It is
normalizable if $l>K/2$, the normalization constant $\alpha_{K1lm}$
depends on the number of the positions $K$, \textit{i.e.} on the
dimension of the amplitude vector $r$ and on $l$ and $m$, see
Sec.~\ref{sec1a1}. Not all moments of this distribution exist, and the
convergence has to be guaranteed when doing integrals. This imposes
conditions on the parameters, we come back to this point.  We notice
that both distributions depend on the quadratic form $r^\dagger
C^{-1}_\textrm{ep} r$ which is known as the Mahalanobis
distance~\cite{Mahalanobis36}.  Importantly, the distribution
\eqref{eq:DetVer} converges to the Gaussian \eqref{eq:GauVer}, when
both parameters $l$ and $m$ are taken to infinity under the condition
\begin{align}
\lim_{l,m\to\infty} \frac{m}{l} &= 2 \ .
\label{eq:GauVerR1}
\end{align}
In the model, the expectation value $\langle rr^\dagger\rangle$ serves
as an estimator for the sample covariances or correlations. We find
\begin{align}
\langle rr^\dagger\rangle_Y &= \beta_Y C_\textrm{ep}
\label{eq:GauVerM1}
\end{align}
with
\begin{align}
 \beta_Y  &= \begin{cases}
                  \displaystyle
                    1               \ ,      & \text{if } \ Y=G  \\
                  \displaystyle
                  \frac{m}{2l-K-2} \ ,      & \text{if } \ Y=A
             \end{cases} \ .
\label{eq:GauVerR1beta}
\end{align}
Due to its very definition, we have $\beta_G=1$ for the multivariate
Gaussian, but a different value $\beta_A$ for the algebraic
distribution. The relation in the latter case then suggests the useful
fixing
\begin{align}
m &= 2l - K - 2 
\label{eq:GauVerM2}
\end{align}
also for finite values of $l$ and $m$. Only with this choice,
$C_\textrm{ep}$ can be estimated by the sample correlation matrix,
otherwise only up to some factor. The distributions \eqref{eq:GauVer}
and \eqref{eq:DetVer} are not sensitive to non--Markovian effects,
even if they are in the data. This is so, because, in a data analysis,
they are obtained by first sampling all amplitudes at all
$T_\textrm{ep}$ times, and then lumping together all these
$T_\textrm{ep}$ distributions. Similarly, one could also consider a
distribution $\widetilde{p}(\widetilde{r}|D_\textrm{ep})$ of the
position amplitudes
$\widetilde{r}=(\widetilde{r}(1),\ldots,\widetilde{r}(T_\textrm{ep}))$,
which is tested by sampling for all $K$ positions, and then lumping
together.  Such a distribution is only sensitive to non--Markovian
effects, but not to the correlations of the time series in the usual
sense.

Both types of correlations, of time and position series, are accounted
for in the distributions that we now choose to model the truly
existing ensemble of correlation matrices, fluctuating around the
positive definite matrices $C$ and $D$. They generalize the amplitude distributions
\eqref{eq:GauVer} and \eqref{eq:DetVer} from $N=1$ to arbitrary $N$.
A natural choice for the distribution of the random data matrix $X$ is
the multimultivariate Gaussian
\begin{align}
  w_G (X|C,D) &= \frac{1}{\sqrt{\det 2\pi D\otimes C}}\exp\left(-\frac{1}{2}\tr D^{-1} X^\dagger C^{-1}X \right) \ ,
  \label{eq:GauVerX}
\end{align}
where we define the direct product $D \otimes C$ as the $N\times N$
matrix with the $K\times K$ matrix entries $D_{nm}C$. The distribution
\eqref{eq:GauVerX} is known in the literature as doubly correlated
Wishart distribution
\cite{Simon2004,Burda2005,McKay2007,Waltner2014,Burda2015}. In
Refs.~\cite{Schmitt2013,Meudt2015} we only considered the Markovian
case, \textit{i.e.}, $D=\mathds{1}_N$, here we go beyond that by
allowing arbitrary positive non--Markovian correlation matrices
$D$. To take care of heavy tails in the distribution of the truly
existing ensemble, we also choose the algebraic, determinant
distribution
\begin{align}
  w_A (X|C,D) &= \frac{\alpha_{KNLM}}{\displaystyle{\det}^L \left( \mathds{1}_N +\frac{1}{M}D^{-1} X^\dagger C^{-1}X \right)} \nonumber\\
   \alpha_{KNLM} &= \sqrt{\frac{2}{M}}^{KN} \prod_{n=1}^{N} \frac{\Gamma(L-(n-1)/2))}{\Gamma(L-(K+n-1)/2)}
                                            \, \frac{1}{\sqrt{\det 2\pi D\otimes C}}
  \label{eq:DetVerX}
\end{align}
depending on two shape parameters $L$ and $M$. This distribution is a
generalization of the ones introduced by Forrester and
Krishnapur~\cite{FK2009} and by Wirtz, Waltner, Kieburg and
Kumar~\cite{WWKK2016}.  Existence of the normalization integral is
guaranteed if
\begin{align}
  L &> \frac{K+N-1}{2} \ .
  \label{eq:DetVerXnorm}
\end{align}
Formally, the amplitude distribution~\eqref{eq:DetVer} is a special
case for $N=1$ and $D=1$. The normalization constant $\alpha_{KNLM}$
includes the normalization constant $\alpha_{K1lm}$ for $N=1$, its
derivation is sketched in Sec.~\ref{sec1a1}. Accordingly, the
distribution \eqref{eq:DetVerX} converges to the Gaussian
\eqref{eq:GauVerX} if $L$ and $M$ are taken to infinity under the
condition
\begin{align}
\lim_{L,M\to\infty} \frac{M}{L} &= 2 \ .
\label{eq:GauVerR2}
\end{align}
The first matrix moment is given by
\begin{align}
  \left\langle \frac{1}{N}XX^\dagger\right\rangle_Y = B_Y C \ ,
\label{eq:GauVerXm}
\end{align}
where
\begin{align}
B_Y  &= \begin{cases}
                  \displaystyle
                    1                 \ ,      & \text{if } \ Y=G  \\
                  \displaystyle
                    \frac{M}{2L-K-N-1}\ ,      & \text{if } \ Y=A
             \end{cases} \ .
\label{eq:GauVerR1Beta}
\end{align}
In the Gaussian case, the confirmation of the value $B_G=1$ is
natural, see Ref.~\cite{Waltner2014}. However, in the algebraic case,
the first marix moment only exists if
\begin{align}
  L &> \frac{K+N+1}{2} \ .
  \label{eq:DetVerXmomex}
\end{align}
As to be expected, the lower threshold is larger by one as in the
condition~\eqref{eq:DetVerXmomex} for the existence of the
normalization. Of course, the value $B_A$ is different from $B_G=1$.
Because the matrix moments in the algebraic case are also of interest
in a general methodical context, we relegate the calculation of the
first and the second ones to a forthcoming study~\cite{GS2020b}.  To
ensure the above mentioned converegence of the algebraic distribution
to the Gaussian, we use the first matrix moment and fix the relation
between the parameters as
\begin{align}
	M &= 2L-K-N-1 \ .
  \label{eq:DetVerXm2}
\end{align}
Only by such a fixing, which puts $B_A$ to one, we guarantee that the
first matrix moment of the distribution~\eqref{eq:DetVerX} is equal to
$C$, facilitating the comparison of tails with the Gaussian case. If
instead of the correlation matrices $C$ and $D$ the covariance
matrices $\Sigma$ and $\Xi$ are employed, we have to replace $C$ in
the relation~\eqref{eq:GauVerXm} with $\Sigma$ and to multiply its
right hand side with $\tr\Xi/N$~\cite{Waltner2014,GS2020b}.

Two observations are worth mentioning. First, the positive
definiteness of $C$ and $D$ allows us to write the matrix in the
trace and in the determinant of Eqs.~\eqref{eq:GauVerX}
and~\eqref{eq:DetVerX} in real--symmetric, positive semidefinite form,
\begin{align}
  D^{-1} X^\dagger C^{-1}X \ &\longrightarrow \ (D^{-1/2} X^\dagger C^{-1/2})(C^{-1/2}XD^{-1/2}) \\
                          & \qquad\qquad = (C^{-1/2}XD^{-1/2})^\dagger (C^{-1/2}XD^{-1/2}) \ .
\label{reals}
\end{align}
We will use this in some of the calculations to follow, we then write,
for $Y=A,G$, $w_Y(\widetilde{X}^\dagger \widetilde{X})$ with
$\widetilde{X}=C^{-1/2}XD^{-1/2}$ instead of $w_Y(X|C,D)$. As the
distributions of the random data matrices only involve matrix invariants,
we also have the remarkable identities
\begin{align}
w_Y(\widetilde{X}^\dagger \widetilde{X}) &= w_Y(\widetilde{X}\widetilde{X}^\dagger) \ , \quad Y=G,A \ ,
\label{reals2}
\end{align}
where the determinant in the algebraic case $Y=A$ is $N \times N$ on
the left--hand and $K \times K$ on the right--hand side.

Second, as the correlation matrices $C_\textrm{ep}$ are dimensionless,
the amplitudes $r$ in the distributions \eqref{eq:GauVer} and
\eqref{eq:DetVer} must be dimensionless as well. If necessary, one can
quickly go over to the distributions involving the covariance matrix
$\Sigma_\textrm{ep}=\sigma_\textrm{ep}C_\textrm{ep}\sigma_\textrm{ep}$
with the variances $\sigma_\textrm{ep}$ as in Eq.~\eqref{corrtm}, we
have
\begin{align}
p_Y(r'|\Sigma_\textrm{ep})\textrm{d}[r'] &= p_Y(r|C_\textrm{ep})\textrm{d}[r]
\label{eq:CS}
\end{align}
for $Y=G,A$ with the new amplitudes $r'=\sigma_\textrm{ep}r$ carrying
a physical dimension. Similarly, we have
\begin{align}
w_Y (X'|\Sigma,\Xi)\textrm{d}[X'] &= w_Y (X|C,D)\textrm{d}[X]
\label{eq:wCS}
\end{align}
with the rescaled, dimension carrying random matrices $X'=\sigma X\xi$.

\section{Methods}
\label{sec1a}

Although the calculations to be carried out in Sec.~\ref{sec2} are
based on a straightforward strategy, the details become rather
involved and complex. To still render the derivations transparent, we
present the calculations in a modular fashion by collecting the
crucial methods and recurring steps here.  In Sec.~\ref{sec1a1}, we
assemble useful representations for the chosen distributions and
present tools for the calculations.  As different conventions can
compel annoying and time consuming checks of formulae, we summarize,
for the convenience of the reader, in Sec.~\ref{sec1a2} the various
special functions and their relations as they are needed in the
sequel.

\subsection{Integral Representations for the Distributions}
\label{sec1a1}

Facilitating the computations to follow, we notice that the algebraic
distribution \eqref{eq:DetVer} is the integral transform
\begin{align}
  p_A(r|C_\textrm{ep}) &= \inte_0^\infty \chi_{2(l-K/2)}^2(z)
                          p_G\left(r\Big|\frac{m}{z}C_\textrm{ep}\right) \textrm{d}z
\label{eq:GauDet}
\end{align}
of the multivariate Gaussian \eqref{eq:GauVer}, involving the $\chi^2$ distribution
\begin{align}
\chi_q^2(z) &= \frac{1}{2^{q/2}\Gamma(q/2)} z^{q/2-1} \exp(-z/2) \Theta(z)
\label{chi2}
\end{align}
of $q$ degrees of freedom. Formula~\eqref{eq:GauDet} has a
generalization to the space of $N\times N$ real--symmetric, positive
definite matrices $Z$, indicated by the notation $Z>0$. We have the
integral transform
\begin{align}
  w_A (\widetilde{X}^\dagger \widetilde{X}) &= 
  \sqrt{\frac{2}{M}}^{KN} \prod_{n=1}^{N} \frac{\Gamma(L-(n-1)/2))}{\Gamma(L-(K+n-1)/2)}
  \int i_{NL}(Z)  w_G \left(\frac{2}{M}Z\widetilde{X}^\dagger\widetilde{X}\right) \textrm{d}[Z] 
  \label{eq:GauDetX}
\end{align}
with $\widetilde{X}=C^{-1/2}XD^{-1/2}$. The volume element
$\textrm{d}[Z]$ is the product of the differentials of all independent
variables. We introduce the Ingham--Siegel distribution
\begin{align}
i_{Nq}(Z) &= \frac{1}{\pi^{N(N-1)/4}\prod_{n=1}^N \Gamma(q-(n-1)/2)} {\det}^{{q-(N+1)/2}} Z \exp(-\tr Z) \Theta(Z)
\label{isd}
\end{align}
which is, apart from a scaling factor in the exponent, a matrix
generalization of the $\chi^2$ distribution~\eqref{chi2}. The matrix
Heaviside distribution $\Theta$ is one, whenever all eigenvalues of
the real--symmetric matrix in its argument are positive and zero
otherwise. Formula~\eqref{eq:GauDetX} is an application of the
Ingham--Siegel integral~\cite{Siegel_1935,Fyodorov_2002}
\begin{align}
  \int_{Z > 0} \exp(-\tr ZR) {\det}^{{q-(N+1)/2}} Z \textrm{d}[Z]
      &= \frac{\pi^{N(N-1)/4}\prod_{n=1}^{N}\Gamma(q-(n-1)/2)}{\det ^q R} \ ,
\label{eq:IngSieInt}
\end{align}
where the matrix $R$ is real--symmetric as well. Convergence is
guaranteed if $q\ge (N+1)/2$. Using Eq.~\eqref{eq:IngSieInt} and the
normalization of the distribution~\eqref{eq:GauVerX}, one easily
derives the normalization constant $\alpha_{KNLM}$ in the
distribution~\eqref{eq:DetVerX}. Owing to Eq.~\eqref{reals2}, we may
express the algebraic distribution of the random data matrix in the
alternative form
\begin{align}
  w_A (\widetilde{X}\widetilde{X}^\dagger) &= 
  \sqrt{\frac{2}{M}}^{KN} \prod_{k=1}^{K} \frac{\Gamma(L-(k-1)/2))}{\Gamma(L-(N+k-1)/2)}
  \int i_{KL}(Z)  w_G \left(\frac{2}{M}Z\widetilde{X}\widetilde{X}^\dagger\right) \textrm{d}[Z] \ ,
  \label{eq:GauDetX2}
\end{align}
where $Z$ is now a positive $K \times K$ integration matrix.

Integrals over the multimultivariate Gaussian~\eqref{eq:GauVerX} and
related expressions are conveniently done as integrals over
multivariate Gaussian--like functions by observing that the trace in
the exponent may, by virtue of
\begin{align}
\tr FX^\dagger GX &= x^\dagger (F^\dagger \otimes G) x \ ,
\label{qf}
\end{align}
be written as quadratic form of the $KN$ component vector $x$,
constructed from the columns $X(n), \ n=1,\ldots,N$, of the $K\times
N$ matrix $X$ according to
\begin{align}
x &= \begin{bmatrix}
               X(1)\\
               \vdots \\
               X(N) 
     \end{bmatrix} \ .
\label{qfv}
\end{align}
Formula~\eqref{qf} holds for real matrices $F$ and $G$ of dimensions
$N\times N$ and $K\times K$, respectively.

To carry out the ensemble averages, we always work with the Fourier
transform, \textit{i.e.} with the charateristic function of the
amplitude distribution, depending on the $K$ component vector $\omega$,
\begin{align}
  \langle \varphi\rangle (\omega|C,D)  &=
  \int \langle p\rangle (r|C,D)  \exp(\text{i} \omega\cdot r) \textrm{d}[r] \nonumber\\
  &= \int \varphi\left(\omega\Big|\frac{1}{N} XX^\dagger\right) w(X|C,D) \textrm{d}[X] \ ,
\label{r3gen}
\end{align}
where we used Eq.~\eqref{rmteac}. Here, $\varphi$ is the
characteristic function of $p$. The corresponding Fourier
backtransform is
\begin{align}
  \langle p\rangle (r|C,D)  &=
  \frac{1}{(2\pi)^K} \int \langle \varphi\rangle (\omega|C,D) \exp(- \text{i} \omega\cdot r) \textrm{d}[\omega] \ .
\label{r3genb}
\end{align}
In view of relation \eqref{eq:GauDet}, it suffices to employ the
Gaussian amplitude distributions,
\begin{align}
  p_G\left(r\Big|\frac{1}{N} XX^\dagger\right) &= \frac{1}{(2\pi)^K} \int
  \varphi_G\left(\omega\Big|\frac{1}{N} XX^\dagger\right)
  \exp(- \text{i} \omega\cdot r) \textrm{d}[\omega] \nonumber\\
  \varphi_G\left(\omega\Big|\frac{1}{N} XX^\dagger\right)
  &= \exp\left(-\frac{1}{2N}\omega^\dagger XX^\dagger\omega\right) \ .
\label{r1}
\end{align}
While the inverse of $XX^\dagger$ appears in the exponent of the
distribution, it is $XX^\dagger$ itself in the charateristic function,
implying that the ensemble average in Eq.~\eqref{r3gen} can, by using
all above results, be reduced to Gaussian integrals.

\subsection{Special Functions Occuring in the Calculations}
\label{sec1a2}

All results on special functions to be summarized here as well as
notations and conventions stem from Ref.~\cite{Gradshteyn2007}.  Due
to the invariances of the random matrix ensembles, we often have to
calculate Fourier back transforms~\eqref{r3genb} of characteristic
functions which depend, due to the structure of our model, on $\omega$
only via a quadratic form involving the correlation matrix $C$. To
indicate that, we write occasionally $\langle \varphi\rangle
(\omega^\dagger C\omega)$ instead of $\langle \varphi\rangle
(\omega|C,D)$. Changing variables according to $\omega\rightarrow
C^{1/2}\omega$, we have
\begin{align}
 \frac{1}{(2\pi)^K} \int \langle \varphi\rangle(\omega^\dagger C\omega) \exp(- \text{i} \omega\cdot r) \textrm{d}[\omega]
 & = \frac{1}{(2\pi)^K\sqrt{\det C}} \int \langle \varphi\rangle(\omega^2) \exp(- \text{i} \omega\cdot C^{-1/2}r)
 \textrm{d}[\omega] \ ,
 \label{bt1}
\end{align}
introduce hyperspherical coordinates with radius $\rho=|\omega|$ and
recognize the angular integrals as spherical Bessel function of zeroth
order in $K$ dimensions. More explicitly, we choose $\vartheta$ as the
angle between $\omega$ and $C^{-1/2}r$, find a factor of
$\sin^{K-2}\vartheta$ in the Jacobian and use the integral
\begin{align}
  \inte_0^\pi \exp(-\text{i}z\cos\vartheta) \sin^\mu\vartheta \textrm{d}\vartheta &=
   \sqrt{\pi} \left(\frac{2}{z}\right)^{\mu/2} \Gamma\left(\frac{\mu+1}{2}\right) J_{\mu/2}(z) \ ,
\label{bt2}
\end{align}
where $J_{\mu/2}(z)$ is the Bessel function of the first kind and of
order $\mu/2$. The remaining angular integrals give a group
volume. Collecting everything, we arrive at
\begin{align}
& \frac{1}{(2\pi)^K} \int \langle \varphi\rangle(\omega^\dagger C\omega) \exp(- \text{i} \omega\cdot r) \textrm{d}[\omega] \nonumber\\
& \qquad\qquad\qquad = \frac{1}{\sqrt{\det 2\pi C}} \frac{1}{\sqrt{r^\dagger C^{-1}r}^{(K-2)/2}}
  \inte_0^\infty \langle \varphi\rangle(\rho^2) J_{(K-2)/2}(\rho\sqrt{r^\dagger C^{-1}r})\rho^{K/2} \textrm{d}\rho \ ,
\label{bt3}
\end{align}
which reduces the $K$ dimensional Fourier backtransform to a certain
one--dimensional Hankel transform. Furthermore, we need the confluent
hypergeometric or Kummer function
\begin{align}
  _1F_1(\mu,\nu,-z) &= \frac{2\Gamma(\nu)}{\Gamma(\mu)\sqrt{z}^{\nu-1}} \inte_0^\infty \exp(-t^2) t^{2\mu-\nu} J_{\nu-1}(2\sqrt{z}t) \textrm{d}t
\label{bt4}
\end{align}
as well as the Tricomi function
\begin{align}
U(\mu,\nu,z) &= \frac{1}{\Gamma(\mu)} \inte_0^\infty \exp(-zt) t^{\mu-1} (1+t)^{\nu-\mu-1} \textrm{d}t \ .
\label{bt5}
\end{align}
We will also employ the integral
\begin{align}
  \inte_0^\infty \frac{J_\nu(bt)t^{\nu+1}}{(a^2+t^2)^{\mu+1}} \textrm{d}t &= \frac{a^{\nu-\mu}b^\mu}{2^\mu\Gamma(\mu+1)} K_{\nu-\mu}(ab)
\label{bt6}
\end{align}
with constants $a$ and $b$ which relates the ordinary and the modified Bessel function of the second
kind or Macdonald function
\begin{align}
  K_\mu(w) &= \frac{w^\mu}{2}  \inte_0^\infty\exp\left(-\frac{t}{2}-\frac{w^2}{2t}\right) t^{-\mu-1} \textrm{d}t
\label{bt6a}
\end{align}
of order $\mu$. The latter appears in the integral
\begin{align}
U(\mu,\nu,z) &= \frac{2}{z^{(\nu-1)/2}\Gamma(\mu)\Gamma(\mu-\nu+1)} \inte_0^\infty \exp(-t) t^{\mu-(\nu+1)/2} K_{\nu-1}(2\sqrt{zt}) \textrm{d}t
\label{bt7}
\end{align}
which connects Tricomi and Macdonald function. The Macdonald function
may also be used to simplify the following double integral over two
$\chi^2$ distributions and an arbitrary function $h(zz')$, depending
on the arguments of the former,
\begin{align}
 \inte_0^\infty \textrm{d}z \chi_{2\kappa}^2(z) \inte_0^\infty \textrm{d}z' \chi_{2\lambda}^2(z') h(zz') &=
     \inte_0^\infty \textrm{d}u h(u) \inte_0^\infty \textrm{d}z \chi_{2\kappa}^2(z) \inte_0^\infty \textrm{d}z' \chi_{2\lambda}^2(z') \delta(u-zz') \nonumber\\
    & = \inte_0^\infty \textrm{d}u h(u) \inte_0^\infty \frac{\textrm{d}z}{z}\chi_{2\kappa}^2(z) \chi_{2\lambda}^2\left(\frac{u}{z}\right) \nonumber\\
    & = \frac{2^{1-\kappa-\lambda}}{\Gamma(\kappa)\Gamma(\lambda)}
         \inte_0^\infty h(u) K_{\lambda-\kappa}(\sqrt{u}) u^{(\lambda+\kappa)/2-1} \textrm{d}u
 \label{bt8}
\end{align} 
owing to the integral representation~\eqref{bt6a}. Finally, we need the Gaussian hypergeometric function
\begin{align}
_2F_1(\lambda,\mu,\nu,z) &= \frac{\Gamma(\nu)}{\Gamma(\mu)\Gamma(\nu-\mu)} \inte_0^1 t^{\mu-1} (1-t)^{\nu-\mu-1}(1-tz)^{-\lambda} \textrm{d}t
\label{bt4a}
\end{align}
which is related to the ordinary and modified Bessel functions by the integrals
\begin{align}
  \inte_0^\infty t^{-\lambda} K_\mu(at) J_\nu(bt) \textrm{d}t &= \frac{b^\nu\Gamma((\nu-\lambda+\mu+1)/2)\Gamma((\nu-\lambda-\mu+1)/2)}
                                                                 {2^{\lambda+1}a^{\nu-\lambda+1}\Gamma(\nu+1)} \nonumber\\
                & \qquad\qquad _2F_1\left(\frac{\nu-\lambda+\mu+1}{2},\frac{\nu-\lambda-\mu+1}{2},\nu+1,-\frac{b^2}{a^2}\right)
\label{bt10}
\end{align}
and
\begin{align}
  \inte_0^\infty t^{-\lambda} K_\mu(at) K_\nu(bt) \textrm{d}t &= \frac{b^\nu\Gamma((\nu-\lambda+\mu+1)/2)\Gamma((\nu-\lambda-\mu+1)/2)}
                                                                 {2^{\lambda+2}a^{\nu-\lambda+1}} \nonumber\\
                                                         & \qquad \frac{\Gamma((1-\lambda+\mu-\nu)/2)\Gamma((1-\lambda-\mu-\nu)/2)}
                                                                 {\Gamma(1-\lambda)} \nonumber\\
                & \qquad _2F_1\left(\frac{\nu-\lambda+\mu+1}{2},\frac{\nu-\lambda-\mu+1}{2},1-\lambda,1-\frac{b^2}{a^2}\right)
\label{bt11}
\end{align}
with constants $a$ and $b$.  All formulas given here are valid within
certain parameter ranges which can be found in
Ref.~\cite{Gradshteyn2007}.

\section{Derivations and Results for the Multivariate Ensemble Averaged Amplitude Distributions}
\label{sec2}

After some preparatory remarks in Sec.~\ref{sec21}, we study the cases
Gaussian--Gaussian, Gaussian--Algebraic, Algebraic--Gaussian and
Algebraic--Algebraic in Secs.~\ref{sec22} to~\ref{sec25},
respectively.

\subsection{General Considerations}
\label{sec21}

The amplitude distributions within the epochs depend on the amplitudes
$r$ and a correlation matrix $C_\textrm{ep}$ considered to be fixed
within the epoch. Its fluctuations are modeled by the distributions of
the random data matrix $X$. We calculate the ensemble averages
\begin{align}
\langle p\rangle_{YY'}(r|C,D) &= \int p_Y\left(r\Big|\frac{1}{N} XX^\dagger\right) w_{Y'}(X|C,D) \textrm{d}[X] 
\label{rmteacGA}
\end{align}
as amplitude distribution for the long time interval for all
combinations $Y,Y'=G,A$.

We also compute the estimators for the sample correlations or
covariances, respectively. They are given by
\begin{align}
\langle rr^\dagger \rangle_{YY'} &= \int rr^\dagger \langle p\rangle_{YY'}(r|C,D) \textrm{d}[r] \ .
\label{e1}
\end{align}
Inserting formula~\eqref{rmteacGA} and interchanging the order of integration, we find
with Eqs.~\eqref{eq:GauVerM1} and~\eqref{eq:GauVerXm}
\begin{align}
  \langle rr^\dagger \rangle_{YY'} &= \int \textrm{d}[X] w_{Y'}(X|C,D)
       \int \textrm{d}[r] rr^\dagger p_Y\left(r\Big|\frac{1}{N} XX^\dagger\right) \nonumber\\
       &= \int \textrm{d}[X] w_{Y'}(X|C,D) \beta_Y \frac{1}{N} XX^\dagger \nonumber\\
       &= \beta_Y B_{Y'} C 
\label{e2}
\end{align}
with $\beta_Y$ and $B_{Y'}$ given in Eqs.~\eqref{eq:GauVerR1beta}
and~\eqref{eq:GauVerR1Beta}.  As in Sec.~\ref{sec14}, if instead of
the correlation matrices $C$ and $D$ the covariance matrices $\Sigma$
and $\Xi$ are used, $C$ in relation~\eqref{e2} has to be replaced with
$\Sigma$ and its right hand side must be multplied with $\tr\Xi/N$.

\subsection{Gaussian--Gaussian}
\label{sec22}

Generalizing the results of Ref.~\cite{Schmitt2013}, we include
non--Markovian effects bei considering a non--trivial correlation
matrix $D\neq \mathds{1}_N$. We find for the ensemble averaged
characteristic function
\begin{align}
  \langle \varphi\rangle_{GG} (\omega|C,D) &= \int \varphi_G\left(\omega\Big|\frac{1}{N} XX^\dagger\right)
                                                             w_G(X|C,D)\textrm{d}[X] \nonumber\\
   &= \frac{1}{\sqrt{\det 2\pi D\otimes C}}
                         \int \exp\left(-\frac{1}{2}x^\dagger\left(\mathds{1}_N\otimes \frac{\omega\omega^\dagger}{N}+
                                   D^{-1}\otimes C^{-1}\right)x\right) \textrm{d}[X] \nonumber\\
   &= \frac{1}{\sqrt{\det(\mathds{1}_N\otimes\mathds{1}_K+D\otimes C\omega\omega^\dagger/N)}} \nonumber\\
   &= \frac{1}{\sqrt{\det(\mathds{1}_N+D\omega^\dagger C\omega/N)}} \ ,
\label{r2}
\end{align}
where we drastically reduce the dimension of the determinant in the
last step with the help of Sylvester's theorem. To calculate the
ensemble averaged distribution 
\begin{align}
  \langle p\rangle_{GG} (r|C,D)  &=
  \frac{1}{(2\pi)^K} \int 
  \frac{\exp(- \text{i} \omega\cdot r)}{\sqrt{\det(\mathds{1}_N+D\omega^\dagger C\omega/N)}}
             \textrm{d}[\omega] \ ,
\label{r4}
\end{align}
\textit{i.e.}, the Fourier backtransform \eqref{r3genb}, we may use formula~\eqref{bt3}
and arrive at
\begin{align}
  \langle p\rangle_{GG} (r|C,D)  &=
  \frac{1}{\sqrt{\det 2\pi C}} \frac{1}{\sqrt{r^\dagger C^{-1}r}^{(K-2)/2}}
  \inte_0^\infty
          \frac{J_{(K-2)/2}(\rho\sqrt{r^\dagger C^{-1}r})}{\sqrt{\det(\mathds{1}_N+D\rho^2/N)}}
                   \rho^{K/2} \textrm{d}\rho \ ,
\label{r6}
\end{align}
which depends, due to the invariances of the ensemble, on the
amplitudes only via the quadratic form $r^\dagger C^{-1}r$.
Furthermore, the matrix $D$ enters formula~\eqref{r6} only by its
eigenvalues $\Upsilon=\diag(\Upsilon(1),\ldots,\Upsilon(N))$ such that
$\det(\mathds{1}_N+D\rho^2/N)=\prod_{n=1}^N(1+\Upsilon(n)\rho^2/N)$.  In
the non--Markovian case $D=\mathds{1}_N$, the integral over $\rho$ in
Eq.~\eqref{r6} can be done with the help of Eq.~\eqref{bt6}
\begin{align}
  \langle p\rangle_{GG} (r|C,\mathds{1}_N)  &=
  \frac{1}{2^{N/2-1}\Gamma(N/2)\sqrt{\det 2\pi C/N}} \frac{K_{(K-N)/2}(\sqrt{Nr^\dagger C^{-1}r})}{\sqrt{Nr^\dagger C^{-1}r}^{(K-N)/2}} \ ,
\label{r7}
\end{align}
confirming the result of Ref.~\cite{Schmitt2013}.

For data analysis, the asymptotic behavior for large arguments
$\sqrt{Nr^\dagger C^{-1}r}$ is important. The Macdonald functions has
a well--known exponential decay, which determines the determines the
asymptotics of $\langle p\rangle_{GG} (r|C,\mathds{1}_N)$. There is no
reason to believe that this exponential behavior turns to an algebraic
one for $D\neq \mathds{1}_N$, we present some arguments in~\ref{app1}.

\subsection{Gaussian--Algebraic}
\label{sec23}

To calculate the ensemble average of the characteristic function, we
employ the integral representation~\eqref{eq:GauDetX} of the algebraic
distribution, interchange the order of the matrix intergals and find
\begin{align}
  \langle \varphi\rangle_{GA} (\omega|C,D) &= \int \varphi_G\left(\omega\Big|\frac{1}{N} XX^\dagger\right)
                                                             w_A(X|C,D)\textrm{d}[X] \nonumber\\
   &= \sqrt{\frac{2}{M}}^{KN} \prod_{n=1}^{N} \frac{\Gamma(L-(n-1)/2))}{\Gamma(L-(K+n-1)/2)}
                                                             \int \textrm{d}[Z] i_{NL}(Z) \nonumber\\
   &  \qquad\qquad \int \varphi_G\left(\omega\Big|\frac{1}{N} XX^\dagger\right)
        w_G \left(\frac{2}{M}Z\widetilde{X}^\dagger\widetilde{X}\right) \textrm{d}[X] \ .
\label{ga1}
\end{align}
Using Eq.~\eqref{qf}, we write the $X$ dependent terms in the exponent as
\begin{align}
  & -\frac{1}{2N}\omega^\dagger XX^\dagger\omega - \frac{1}{M}\tr Z\widetilde{X}^\dagger\widetilde{X} \nonumber\\
  & \qquad\qquad\qquad = -\frac{1}{2N}\tr X^\dagger\omega\omega^\dagger X - \frac{1}{M}\tr D^{-1/2}ZD^{-1/2}X^\dagger CX \nonumber\\
  & \qquad\qquad\qquad = -\frac{1}{2}x^\dagger \left(\mathds{1}_N\otimes \frac{\omega\omega^\dagger}{N}+
                                     \frac{2}{M}D^{-1/2}ZD^{-1/2} \otimes C^{-1}\right)x
  \label{ga2}
\end{align}
and do the Gaussian integral over $X$ which yields
\begin{align}
& \frac{\sqrt{2\pi}^{KN}}{\sqrt{\det\left(\mathds{1}_N\otimes \omega\omega^\dagger/N+
                                            2D^{-1/2}ZD^{-1/2}/M \otimes C^{-1}\right)}} \nonumber\\
& \qquad = \frac{\sqrt{2\pi}^{KN}{\det}^{N/2}C}{{\det}^{(K-1)/2}(2D^{-1/2}ZD^{-1/2}/M)\sqrt{\det(\omega^\dagger C\omega\mathds{1}_N/N+2D^{-1/2}ZD^{-1/2}/M)}} \nonumber\\
& \qquad = \sqrt{\frac{M}{2}}^{N(K-1)} \frac{\sqrt{2\pi}^{KN}\sqrt{\det D\otimes C}}{{\det}^{(K-1)/2}Z\sqrt{\det(\omega^\dagger C\omega D/N+2Z/M)}} \nonumber\\
  & \qquad = \frac{\sqrt{\pi M}^{N(K-1)}\sqrt{\det D\otimes C}}{{\det}^{(K-1)/2}Z}
                 \int \exp\left(-\zeta^\dagger\left(\frac{\omega^\dagger C\omega}{2N}D +\frac{1}{M}Z\right)\zeta\right) \textrm{d}[\zeta] \ .
\label{ga3}
\end{align}
Again, it was possible to considerably reduce the dimension of the
determinant, the resulting determinant is written as a Gaussian
integral over an $N$ component vector $\zeta$. Inserting this into
Eq.~\eqref{ga1}, we have
\begin{align}
 \langle \varphi\rangle_{GA} (\omega|C,D) &= \frac{1}{\sqrt{\pi M}^N} \prod_{n=1}^{N} \frac{\Gamma(L-(n-1)/2)}{\Gamma(L-(K+n-1)/2)}
            \int \textrm{d}[Z] \frac{i_{NL}(Z)}{{\det}^{(K-1)/2}Z} \nonumber\\
     & \qquad\qquad \int \exp\left(-\zeta^\dagger\left(\frac{\omega^\dagger C\omega}{2N}D +\frac{1}{M}Z\right)\zeta\right) \textrm{d}[\zeta] \ . 
\label{ga4}
\end{align}
Observing $\zeta^\dagger Z\zeta=\tr \zeta\zeta^\dagger Z$, the matrix integral is conveniently
recognized as an Ingham--Siegel integral~\eqref{eq:IngSieInt},
\begin{align}
&  \int \frac{i_{NL}(Z)}{{\det}^{(K-1)/2}Z} \exp\left(-\zeta^\dagger \frac{1}{M}Z \zeta\right) \textrm{d}[Z]  \nonumber\\
   & \qquad\qquad = \prod_{n=1}^{N} \frac{\Gamma(L-(K-1)/2-(n-1)/2)}{\Gamma(L-(n-1)/2)}\frac{1}{{\det}^{L-(K-1)/2} (\mathds{1}_N+\zeta\zeta^\dagger/M)} \ .
\label{ga5}
\end{align}
Once more, we can simplify the determinant, 
\begin{align}
  \frac{1}{{\det}^{L-(K-1)/2} (\mathds{1}_N+\zeta\zeta^\dagger/M)} &= \frac{1}{(1+\zeta^\dagger\zeta/M)^{L-(K-1)/2}}  \nonumber\\
    &= \inte_0^\infty \chi_{2(L-(K-1)/2)}^2(z) \exp\left(-\frac{\zeta^\dagger\zeta}{2M}z\right) \textrm{d}z \ ,
\label{ga6}
\end{align}
where we used Eqs.~\eqref{eq:GauDet} and~\eqref{chi2}. Many of the remaining $\Gamma$ functions cancel.
Collecting everything we observe that the $\zeta$ integral can be done and
\begin{align}
 \langle \varphi\rangle_{GA} (\omega|C,D) &= \sqrt{\frac{2}{M}}^N \frac{\Gamma(L-(K-1)/2)}{\Gamma(L-(K+N-1)/2)}
 \inte_0^\infty \frac{\chi_{2(L-(K-1)/2)}^2(z)\textrm{d}z}{\sqrt{\det(z\mathds{1}_N/M+\omega^\dagger C\omega D/N)}}
 \label{ga7}
\end{align} 
is the final expression for the ensemble averaged characteristic
function. By virtue of Eq.~\eqref{bt3}, we find
\begin{align}
\langle p\rangle_{GA} (r|C,D) &= \sqrt{\frac{2}{M}}^N \frac{\Gamma(L-(K-1)/2)}{\Gamma(L-(K+N-1)/2)}
  \frac{1}{\sqrt{\det 2\pi C}\sqrt{r^\dagger C^{-1}r}^{(K-2)/2}} \nonumber\\
  & \qquad\qquad \inte_0^\infty \inte_0^\infty 
  \frac{J_{(K-2)/2}(\rho\sqrt{r^\dagger C^{-1}r})\chi_{2(L-(K-1)/2)}^2(z)}{\sqrt{\det(z\mathds{1}_N/M+\rho^2 D/N)}}
  \textrm{d}z \rho^{K/2} \textrm{d}\rho \ ,
 \label{ga8}
\end{align} 
which expresses the ensemble averaged amplitude distribution as
twofold integral. As pointed out in Sec.~\ref{sec14}, the algebraic
distribution $w_A(X|C,D)$ converges to the Gaussian $w_G(X|C,D)$ in
the limit $L,M\to\infty$ under the
condition~\eqref{eq:GauVerR2}. Consequently, this limit of
Eq.~\eqref{ga8} also reproduces the result~\eqref{r6}. This is easily
shown by rescaling $z\rightarrow zM$ and then carrying out a
saddlepoint approximation for the $z$ integral, the saddlepoint lies
at $z=1$.

One of the integrals in Eq.~\eqref{ga8} can be carried out by making
the change of variables $z\rightarrow\rho^2z$ which removes the $\rho$
dependence from the determinant. The $\rho$ integral is then of the
form \eqref{bt4}, yielding the Kummer function. Changing variables
according to $z=M/Nu$ we eventually end up with
\begin{align}
  \langle p\rangle_{GA} (r|C,D) &= \frac{\Gamma(L-(N-1)/2)}{\Gamma(L-(K+N-1)/2)\Gamma(K/2)}
  \frac{1}{\sqrt{\det2\pi CM/N }} \nonumber\\
  & \qquad \inte_0^\infty {_1F_1}\left(L-\frac{N-1}{2},\frac{K}{2},-\frac{uN}{2M}r^\dagger C^{-1} r\right)
  \frac{u ^{(K-2)/2}\textrm{d}u}{\sqrt{\det \left(\mathds{1}_N +uD\right)}} \ .
\label{ga9}
\end{align}
In the Markovian case $D=\mathds{1}_N$, the determinant becomes a
power and formula \eqref{bt5} allows us to do the integral,
\begin{align}
  \langle p\rangle_{GA} (r|C,\mathds{1}_N) &= \frac{\Gamma(L-(N-1)/2)\Gamma(L-(K-1)/2)}{\Gamma(L-(K+N-1)/2)\Gamma(N/2)}
  \frac{1}{\sqrt{\det2\pi CM/N }} \nonumber\\
  & \qquad\qquad\qquad U\left(L-\frac{N-1}{2},\frac{K-N}{2}+1,\frac{N}{2M}r^\dagger C^{-1} r\right) \ ,
\label{ga10}
\end{align}
which is a closed form expression involving the Tricomi function.

Importantly, the asymptotic behavior of $\langle p\rangle_{GA}
(r|C,D)$ is algebraic, not exponential as in the case of a Gaussian
distribution of the random data matrices. More precisely, we have
\begin{align}
  \langle p\rangle_{GA} (r|C,D) &\sim \frac{1}{(r^\dagger C^{-1}
    r)^{L-(N-1)/2}}
\label{ga11}
\end{align}
for large values of $\sqrt{r^\dagger C^{-1} r}$. Of course,
this behavior comes from the algebraic distribution of the random
data matrices, the derivation is given in~\ref{app1}.

\subsection{Algebraic--Gaussian}
\label{sec24}

Major steps in the calculation can be traced back to the
Gaussian--Gaussian case. Using Eq.~\eqref{eq:GauDet} in
Eq.~\eqref{rmteacGA} and interchanging integrals, we find
\begin{align}
  \langle p\rangle_{AG}(r|C,D) &= \inte_0^\infty \textrm{d}z \chi_{2(l-K/2)}^2(z)
  \int p_G\left(r\Big|\frac{m}{z}\frac{1}{N} XX^\dagger\right) w_G(X|C,D) \textrm{d}[X] \nonumber\\
  &= \inte_0^\infty \textrm{d}z \chi_{2(l-K/2)}^2(z) \sqrt{\frac{z}{m}}^K
  \int p_G\left(\sqrt{\frac{z}{m}}r\Big|\frac{1}{N} XX^\dagger\right) w_G(X|C,D) \textrm{d}[X]  \nonumber\\
  &= \inte_0^\infty \chi_{2(l-K/2)}^2(z) \sqrt{\frac{z}{m}}^K
                        \langle p\rangle_{GG}\left(\sqrt{\frac{z}{m}}r\Big|C,D\right) \textrm{d}z 
\label{ag1}
\end{align}
for the ensemble averaged amplitude distribution and
\begin{align}
  \langle\varphi\rangle_{AG}(\omega|C,D) &= \inte_0^\infty \chi_{2(l-K/2)}^2(z) 
                        \langle \varphi\rangle_{GG}\left(\sqrt{\frac{m}{z}}\omega\Big|C,D\right) \textrm{d}z 
\label{ag2}
\end{align}
for the characteristic function, which both are one--dimensional
$\chi^2$ transforms of the corresponding function in the
Gaussian--Gaussian case. Inserting formula \eqref{r2} we arrive at
\begin{align}
 \langle \varphi\rangle_{AG} (\omega|C,D) &= \sqrt{\frac{2}{m}}^N \frac{\Gamma(l-(K-N)/2)}{\Gamma(l-K/2)}
 \inte_0^\infty \frac{\chi_{2(l-(K-N)/2)}^2(z)\textrm{d}z}{\sqrt{\det(z\mathds{1}_N/m+\omega^\dagger C\omega D/N)}}
 \label{ag3}
\end{align} 
for the characteristic function, which coincides with the
result~\eqref{ga7} in the Gaussian--Algebraic case if we replace $L$,
$M$ and $N$ with proper combinations of $l$, $m$ and $N$, which are,
however, not easy to guess as the different dependencies on these
parameters have different origin. With formula~\eqref{bt3}, we find
\begin{align}
\langle p\rangle_{AG} (r|C,D) &= \sqrt{\frac{2}{m}}^N \frac{\Gamma(l-(K-N)/2)}{\Gamma(l-K/2)}
  \frac{1}{\sqrt{\det 2\pi C}\sqrt{r^\dagger C^{-1}r}^{(K-2)/2}} \nonumber\\
  & \qquad\qquad \inte_0^\infty \inte_0^\infty 
  \frac{J_{(K-2)/2}(\rho\sqrt{r^\dagger C^{-1}r})\chi_{2(l-(K-N)/2)}^2(z)}{\sqrt{\det(z\mathds{1}_N/m+\rho^2 D/N)}}
  \textrm{d}z \rho^{K/2} \textrm{d}\rho
 \label{ag4}
\end{align} 
for the ensemble averaged amplitude distribution. It also coincides
mathematically with the result \eqref{ga8} in the Gaussian--Algebraic
case if proper, but non--trivial parameter replacements are made.
With steps as in Sec.~\ref{sec23}, we can do one of the integrals in
Eq.~\eqref{ag4} and find
\begin{align}
  \langle p\rangle_{AG} (r|C,D) &= \frac{\Gamma(l)}{\Gamma(l-K/2)\Gamma(K/2)}
  \frac{1}{\sqrt{\det2\pi Cm/N }} \nonumber\\
  & \qquad \inte_0^\infty {_1F_1}\left(l,\frac{K}{2},-\frac{uN}{2m}r^\dagger C^{-1} r\right)
  \frac{u ^{(K-2)/2}\textrm{d}u}{\sqrt{\det \left(\mathds{1}_N +uD\right)}} 
\label{ag5}
\end{align}
for arbitrary $D$ as well as
\begin{align}
  \langle p\rangle_{AG} (r|C,\mathds{1}_N) &= \frac{\Gamma(l)\Gamma(l-(K-N)/2)}{\Gamma(l-K/2)\Gamma(N/2)}
  \frac{1}{\sqrt{\det2\pi Cm/N }} \nonumber\\
  & \qquad\qquad\qquad U\left(l,\frac{K-N}{2}+1,\frac{N}{2m}r^\dagger C^{-1} r\right)
\label{ag6}
\end{align}
in the Markovian case $D=\mathds{1}_N$.  The similarity of the
distributions to those given in Sec.~\ref{sec23} does not come as a
surprise, but it is a purely mathematical one. From the viewpoint of
physics and data analysis, the cases Gaussian--Algebraic and
Algebraic--Gaussian are very different and only identical for the
irrelevant parameter value $N=1$, corresponding to model time series
of length one.

For arbitrary $D$, the asymptotic behavior of $\langle
p\rangle_{AG}(r|C,D)$ can be infered from Sec.~\ref{sec23} and~\ref{app1}. We find
\begin{align}
  \langle p\rangle_{AG} (r|C,D) &\sim \frac{1}{(r^\dagger C^{-1} r)^l}
\label{ag7}
\end{align}
for large values of $\sqrt{r^\dagger C^{-1} r}$, which results from
the asymptotic relation~\eqref{ga11} by replacing $L-(N-1)/2$ with $l$.
This is plausibel, because the algebraic behavior stems only from the
amplitude distributions within the epochs which formally coincides
with the distribution of the random data matrices for $l=L$ and $N=1$.

\subsection{Algebraic--Algebraic}
\label{sec25}

As in the Algebraic--Gaussian case we use a shortcut, here by
observing that the desired functions can be written as integrals over
the corresponding ones in the Gaussian--Algebraic case. The steps in
Eq.~\eqref{ag1} carry over to an algebraic distribution of the random
data matrix, yielding for the ensemble averaged amplitude distribution
the integral
\begin{align}
  \langle p\rangle_{AA}(r|C,D) &= \inte_0^\infty \chi_{2(l-K/2)}^2(z)
  \sqrt{\frac{z}{m}}^K \langle p\rangle_{GA}\left(\sqrt{\frac{z}{m}}r\Big|C,D\right) \textrm{d}z
\label{aa1}
\end{align}
with $\langle p\rangle_{GA}$ calculated in Sec.~\ref{sec23}. Similarly,
we have
\begin{align}
  \langle\varphi\rangle_{AA}(\omega|C,D) &= \inte_0^\infty \chi_{2(l-K/2)}^2(z) 
                        \langle \varphi\rangle_{GA}\left(\sqrt{\frac{m}{z}}\omega\Big|C,D\right) \textrm{d}z 
\label{aa2}
\end{align}
for the characteristic function. Pluging in the result~\eqref{ga7} and
rearranging terms, we find for the latter
\begin{align}
  \langle\varphi\rangle_{AA}(\omega|C,D) &= 
  \frac{2^N}{\sqrt{Mm}^N} \frac{\Gamma(l-(K-N)/2)\Gamma(L-(K-1)/2)}{\Gamma(l-K/2)\Gamma(L-(K+N-1)/2)} \nonumber\\
& \qquad\qquad  \inte_0^\infty\inte_0^\infty \frac{\chi_{2(L-(K-1)/2)}^2(z)\chi_{2(l-(K-N)/2)}^2(z')\textrm{d}z\textrm{d}z'}
                   {\sqrt{\det(zz'\mathds{1}_N/(Mm)+\omega^\dagger C\omega D/N)}} \ ,
\label{aa3}
\end{align}
which may, by virtue of Eq.~\eqref{bt8}, be cast into the form of a certain one--dimensional
Bessel transform,
\begin{align}
  \langle \varphi\rangle_{AA} (\omega|C,D) &= \frac{2^{K+(N+1)/2-L-l}}{\sqrt{Mm}^N\Gamma(l-K/2)\Gamma(L-(K+N-1)/2)}
                \nonumber\\
& \qquad\qquad \inte_0^\infty \frac{K_{l-L+(N-1)/2}(\sqrt{u})\sqrt{u}^{L+l-K+(N+1)/2-2}}
                     {\sqrt{\det(u\mathds{1}_N/(Mm)+\omega^\dagger C\omega D/N)}} \textrm{d}u \ .
 \label{aa4}
\end{align} 
Applying formula~\eqref{bt3} yields after some algebra the ensemble
averaged amplitude distribution
\begin{align}
  \langle p\rangle_{AA} (r|C,D) &= \frac{2^{K+(N+1)/2-L-l}}{\sqrt{\det 2\pi C}\sqrt{r^\dagger C^{-1}r}^{(K-2)/2}\sqrt{Mm}^N\Gamma(l-K/2)\Gamma(L-(K+N-1)/2)}
                \nonumber\\
& \quad \inte_0^\infty \inte_0^\infty \frac{J_{(K-2)/2}(\rho\sqrt{r^\dagger C^{-1}r})K_{l-L+(N-1)/2}(\sqrt{u})\sqrt{u}^{L+l-K+(N+1)/2-2}\rho^{K/2}}
                     {\sqrt{\det(u\mathds{1}_N/(Mm)+\rho^2D/N)}} \textrm{d}u\textrm{d}\rho \ .
 \label{aa5}
\end{align}
Analogously to Sec.~\ref{sec23}, the $\rho$ integral can be performed
by making the change of variables $u\rightarrow u\rho^2$, which
removes $\rho$ from the determinant. The $\rho$ integral is then of
the form \eqref{bt10} and gives the Gaussian hypergeometric
function. With the change of variables $u=Mm/Nv$, we arrive at
\begin{align}
  \langle p\rangle_{AA} (r|C,D) &= \frac{\Gamma(l)\Gamma(L-(N-1)/2)}{\sqrt{\det\pi CMm/N}\Gamma(l-K/2)\Gamma(L-(K+N-1)/2)\Gamma(K/2)}                  
                  \nonumber\\
                  & \qquad\qquad \inte_0^\infty {_2F_1}\left(l,L-\frac{N-1}{2},\frac{K}{2},-\frac{Nr^\dagger C^{-1}r}{Mm}v\right)
                   \frac{v^{(K-2)/2}\textrm{d}v}{\sqrt{\det(\mathds{1}_N+vD)}} \ .
\label{aa6}
\end{align}
Even in the Algebraic--Algebraic case, a reduction to a
one--dimensional integral can be carried out for arbitrary $D$. In the
Markovian case $D=\mathds{1}_N$, we start from Eq.~\eqref{aa5}, use
formula~\eqref{bt6} for the $\rho$ integral which allows us to apply
formula~\eqref{bt11} to the remaining $u$ integral,
\begin{align}
  \langle p\rangle_{AA} (r|C,\mathds{1}_N) &= \frac{\Gamma(l)\Gamma(l-(K-N)/2)}{\sqrt{\det\pi CMm/N}\Gamma(l-K/2)}  \nonumber\\
            & \qquad\qquad \frac{\Gamma(L-(N-1)/2)\Gamma(L-(K-1)/2)}{\Gamma(L-(K+N-1)/2)\Gamma(L+l-(K-1)/2)\Gamma(N/2)}                  
                  \nonumber\\
                  & \qquad\qquad {_2F_1}\left(l,L-\frac{N-1}{2},L+l-\frac{K-1}{2},1-\frac{Nr^\dagger C^{-1}r}{Mm}\right) \ ,
 \label{aa7}
\end{align}
which is a complicated, but still a closed form expression as in the previous cases.

As both, the amplitude distributions within the epochs with parameter
$l$ as well as the distribution of the random data matrices with
parameters $L$ and $N$, are algebraic, the asymptotic behavior may be
governed by either of them,
\begin{align}
  \langle p\rangle_{AA} (r|C,D) &\sim \begin{cases}
                                      \displaystyle
                                      \frac{1}{(r^\dagger C^{-1}r)^l} \ ,     & \text{if } l<L-\frac{N-1}{2}  \\
                                      \displaystyle
                                      \frac{\ln r^\dagger C^{-1}r}{(r^\dagger C^{-1}r)^l} \ ,      & \text{if } l=L-\frac{N-1}{2}  \\
                                      \displaystyle
                                      \frac{1}{(r^\dagger C^{-1}r)^{L-(N-1)/2}} \ ,     & \text{if } l>L-\frac{N-1}{2} 
                                     \end{cases} \ .
  \label{aa8}
\end{align}  
The derivation is given in~\ref{app1}.

\section{A Technique and Formulae for the Analysis of Data}
\label{sec3}

We begin with general considerations in Sec.~\ref{sec31}, before we
collect the results for the four cases in Secs.~\ref{sec32}
to~\ref{sec35}. In Sec.~\ref{sec36}, some figures illustrate our
results.

\subsection{General Considerations}
\label{sec31}

To compare our $K$ variate distributions with data, we further extend
the method introduced in Ref.~\cite{Schmitt2013} which we developed in
the spirit of aggregation. The crucial idea is to construct $K$
univariate distributions out of the $K$ variate one which are then
overlaid, \textit{i.e.}, lumped together, or, if meaningful, analyzed
individually. We take advantage of the fact that all $K$ variate
distributions depend on the amplitudes $r$ via the Mahalanobis
distance $r^\dagger C^{-1}r$ only. As anticipated in Sec.~\ref{sec1a}
and explicitly shown in Sec.~\ref{sec2}, the Fourier
backtransform~\eqref{r3genb} which yields the ensemble averaged
amplitude distribution $\langle p\rangle_{YY'} (r|C,D)$ has always the
form~\eqref{bt1}. To decouple the amplitudes, we rotate the vektor $r$
into the eigenbasis of the correlation matrix $C$. More precisely, we
use the diagonalization
\begin{align}
  C &= U\Lambda U^\dagger \qquad \text{such that} \qquad   C^{-1/2} = U\Lambda^{-1/2} U^\dagger \ ,
 \label{sec3.1}
\end{align}
where $U$ is an orthogonal $K\times K$ matrix and $\Lambda$ is the
diagonal matrix of the eigenvalues $\Lambda_k$. As they are positive definite,
the square roots $\Lambda_k^{1/2}$ are real, we choose them positive. We use
the rotated amplitudes
\begin{align}
\bar{r} &= U^\dagger r
 \label{sec3.2}
\end{align}
as new arguments of the ensemble averaged amplitude distribution. The
corresponding Jacobi determinant is one and the functional form of the
distribution is thus not altered. In the exponential function on the
right hand side of Eq.~\eqref{bt1}, we have
\begin{align}
  \omega \cdot C^{-1/2}r &= \omega \cdot U\Lambda^{-1/2}\bar{r}
                        = (U^\dagger\omega) \cdot \Lambda^{-1/2}\bar{r} \ .
 \label{sec3.3}
\end{align}
As the characteristic function $\langle \varphi\rangle_{YY'}$ depends on the
vector $\omega$ via its length only, the change $\omega\rightarrow
U^\dagger\omega$ of integration variables fully removes $U$ from the
integrand. We find
\begin{align}
  \langle p\rangle_{YY'} (\bar{r}|C,D)
  & = \frac{1}{(2\pi)^K\sqrt{\det C}} \int \langle \varphi\rangle_{YY'}(\omega^2)
         \exp(- \text{i} \omega\cdot\Lambda^{-1/2}\bar{r}) \textrm{d}[\omega] \ .
\label{sec3.4}
\end{align}
We are now ready to define $K$ univariate distributions by integrating
out all rotated amplitudes but the $k$--th one,
\begin{align}
  \langle p\rangle_{YY'}^{(k)} (\bar{r}_k|C,D)  &= \int \langle p\rangle_{YY'} (\bar{r}|C,D) \textrm{d}[\bar{r}]_{\neq k} \, \qquad k=1,\ldots,K \ .
\label{sec3.5}
\end{align}
Inserting Eq.~\eqref{sec3.4}, we can do all these $K-1$ integrals which give
$\prod_{l\neq k}2\pi\Lambda_l^{1/2}\delta(\omega_l)$. Hence, all $\omega$ integrals
except the $k$--th one can be carried out and we arrive at a one--dimensional Fourier
transform,
\begin{align}
  \langle p\rangle_{YY'}^{(k)} (\bar{r}_k|\Lambda_k,D)  &= \frac{1}{\pi\sqrt{\Lambda_k}}
          \inte_0^\infty \langle \varphi\rangle_{YY'}(\omega_k^2) \cos\frac{\omega_k\bar{r}_k}{\sqrt{\Lambda_k}}\textrm{d}\omega_k \ ,
\label{sec3.6}
\end{align}
which reduces to a Fourier cosine transform because $\langle \varphi\rangle_{YY'}$ is an even
function. We thus obtain $K$ univariate distributions of the same
form, but scaled with $\sqrt{\Lambda_k}$. It is helpful to rewrite
Eq.~\eqref{sec3.6} by using
\begin{align}
\cos z &= \sqrt{\frac{\pi z}{2}}J_{-1/2}(z) 
\label{sec3.7}
\end{align}
which yields
\begin{align}
  \langle p\rangle_{YY'}^{(k)} (\bar{r}_k|\Lambda_k,D)  &=
  \frac{1}{\sqrt{2\pi\Lambda_k}} \left(\frac{\bar{r}_k^2}{\Lambda_k}\right)^{1/4}
   \inte_0^\infty \langle \varphi\rangle_{YY'}(\rho^2)J_{-1/2}\left(\sqrt{\frac{\bar{r}_k^2}{\Lambda_k}}\rho\right) \sqrt{\rho}\textrm{d}\rho \ .
\label{sec3.9}
\end{align}
This formula greatly simplifies the ensuing derivations that involve
algebraic distributions, as we can proceed exactly as in
Sec.~\ref{sec2}, we only have to set $K=1$ and replace $r^\dagger
C^{-1}r$ with $\bar{r}_k^2/\Lambda_k$ in the term
$J_{(K-2)/2}(\rho\sqrt{r^\dagger C^{-1}r})\rho^{K/2}/\sqrt{r^\dagger
  C^{-1}r}^{(K-2)/2}$ related to the spherical Bessel function. We
notice that $\langle\varphi\rangle_{YY'}(\rho^2)$ is unchanged and, in
general, depends on $K$ directly as well as implicitly. Furthermore,
the arguments leading to the asymptotic behavior carry over to the
present discussion.

The variances of the univariate distributions $\langle
p\rangle_{YY'}^{(k)} (\bar{r}_k|\Lambda_k,D)$ follow directly from the
result~\eqref{e2}. Due to the symmetry of the distributions, they
coincide with the second moments $\langle \bar{r}_k^2
\rangle_{YY'}^{(k)}$. Consider the left hand side of Eq.~\eqref{e2} in
the rotated coordinates~\eqref{sec3.2},
\begin{align}
  \langle rr^\dagger \rangle_{YY'} &= U\langle\bar{r} \bar{r}^\dagger\rangle_{YY'} U^\dagger
\label{vk1}
\end{align}
and thus
\begin{align}
  \langle\bar{r} \bar{r}^\dagger\rangle_{YY'} &= U^\dagger \langle rr^\dagger\rangle_{YY'} U
  =  U^\dagger \beta_YB_{Y'} C U =  \beta_YB_{Y'} \Lambda \ .
\label{vk2}
\end{align}
Naturally, the correlation matrix in the eigenbasis of $C$ is
diagonal and proportional to $\Lambda$. This implies
\begin{align}
  \langle \bar{r}_k^2 \rangle_{YY'}^{(k)} &= \beta_Y B_{Y'} \Lambda_k 
\label{sec3.9a}
\end{align}
with $\beta_Y$ and $B_{Y'}$ given in Eqs.~\eqref{eq:GauVerR1beta}
and~\eqref{eq:GauVerR1Beta}.

When analyizing data, $K$ is given, we obtain the matrices $C$ and $D$
or $\Sigma$ and $\Xi$ by using the the originally measured amplitudes
for sampling over the long time interval $T_\textrm{tot}$.  The data
are normalized in different ways for the computation of the time and
the position series, see Sec.~\ref{sec11}, implying that $C$ and $D$
have different eigenvalues.  The result~\eqref{sec3.9a} is important
as $\langle \bar{r}_k^2 \rangle_{YY'}$ serves as sample variance for
the univariate distributions $\langle p\rangle_{YY'}^{(k)}
(\bar{r}_k|\Lambda_k,D)$ in the eigenbasis of the correlation matrix,
\textit{i.e.}  for the rotated amplitudes. This variance allows one to
determine an unknown parameter in the underlying algebraic
distributions, if the fixings~\eqref{eq:GauVerM2}
and~\eqref{eq:DetVerXm2} have not been made. In the
Gaussian--Algebraic and Algebraic--Gaussian cases, the
result~\eqref{sec3.9a} provides relations between the unknown
parameters $M$, $L$, $N$ and $m$, $l$, respectively.  In the
Algebraic--Algebraic case, there is only one relation for all of these
parameters, hence, only one, not both, can be inferred from the sample
variance. The relation between $l$ and $m$ can be obtained within the
epochs by sampling variances or by fitting.  In all cases, the
parameter $N$ is a fit parameter, measuring the strength of the
fluctuations. It might be helpful to use higher moments to obtain
further relations. However, experience tells, that $N$ sensitively
determines the shape and is best obtained by fitting the whole
distribution $\langle p\rangle_{YY'}^{(k)} (\bar{r}_k|\Lambda_k,D)$ to
the data.  Once more, if instead of the correlation matrices $C$ and
$D$ the covariance matrices $\Sigma$ and $\Xi$ are used, $\Lambda_k$
in Eq.~\eqref{sec3.9a} has to be interpreted as the $k$--th eigenvalue
of $\Sigma$ and the right hand side of Eq.~\eqref{sec3.9a} must be
multiplied with $\tr\Xi/N$.

\subsection{Gaussian--Gaussian}
\label{sec32}

In a general non--Markovian case with $D\neq \mathds{1}_N$, we obtain the
one--dimensional integral
\begin{align}
 \langle p\rangle_{GG}^{(k)} (\bar{r}_k|\Lambda_k,D)  &= \frac{1}{\pi\sqrt{\Lambda_k}}
                 \inte_0^\infty \frac{1}{\det(\mathds{1}_N+D\rho^2/N)}
                 \cos\frac{\rho\bar{r}_k}{\sqrt{\Lambda_k}}\textrm{d}\rho
\label{sec3.10}
\end{align}
which reduces in the Markovian situation $D=\mathds{1}_N$ to
\begin{align}
  \langle p\rangle_{GG}^{(k)} (\bar{r}_k|\Lambda_k,\mathds{1}_N) &=
  \frac{1}{2^{(N-1)/2}\Gamma(N/2)\sqrt{\pi \Lambda_k/N}} \sqrt{\frac{N\bar{r}_k^2}{\Lambda_k}}^{(N-1)/2}
                          K_{(1-N)/2}\left(\sqrt{\frac{N\bar{r}_k^2}{\Lambda_k}}\right) 
\label{sec3.11}
\end{align}
in agreement with Ref.~\cite{Schmitt2013}. This case is exceptional,
because its Gaussian nature leads to a very simple relation bewteen
the results~\eqref{sec3.11} and~\eqref{r7}. The former follows from
the latter by just putting $K=1$. This is not so in the other three
cases.

\subsection{Gaussian--Algebraic}
\label{sec33}

Carrying out steps analogous to the ones in Sec.~\ref{sec23}, we
obtain the formulae for a general Markovian case with $D\neq
\mathds{1}_N$,
\begin{align}
  \langle p\rangle_{GA}^{(k)} (\bar{r}_k|\Lambda_k,D) &= \frac{\Gamma(L-(K+N)/2+1)}{\Gamma(L-(K+N-1)/2)\pi\sqrt{2\Lambda_kM/N }} \nonumber\\
  & \qquad \inte_0^\infty {_1F_1}\left(L-\frac{K+N}{2}+1,\frac{1}{2},-\frac{uN}{2M}\frac{\bar{r}_k^2}{\Lambda_k}\right)
  \frac{\textrm{d}u}{\sqrt{u\det \left(\mathds{1}_N +uD\right)}} \ ,
\label{sec3.12}
\end{align}
as well as for a Markovian situation with $D=\mathds{1}_N$,
\begin{align}
  \langle p\rangle_{GA}^{(k)} (\bar{r}_k|\Lambda_k,\mathds{1}_N) &= \frac{\Gamma(L-(K+N)/2+1)\Gamma(L-(K-1)/2)}
               {\Gamma(L-(K+N-1)/2)\Gamma(N/2)\sqrt{2\pi\Lambda_kM/N }}\nonumber\\
  & \qquad\qquad\qquad\qquad 
             U\left(L-\frac{K+N}{2}+1,\frac{1-N}{2}+1,\frac{N}{2M}\frac{\bar{r}_k^2}{\Lambda_k}\right) \ .
\label{sec3.13}
\end{align}
We also find
\begin{align}
  \langle p\rangle_{GA}^{(k)} (\bar{r}_k|\Lambda_k,D) &\sim \left(\frac{\bar{r}_k^2}{\Lambda_k}\right)^{-L+(K+N)/2-1}
\label{sec3.14}
\end{align}
as asymptotic result.

\subsection{Algebraic--Gaussian}
\label{sec34}

This case is, as already mentioned in Sec.~\ref{sec2}, to some extent
similar to the previous one and we arrive for arbitrary $D$ at
\begin{align}
  \langle p\rangle_{AG}^{(k)} (\bar{r}_k|\Lambda_k,D) &= \frac{\Gamma(l-(K-1)/2)}{\Gamma(l-K/2)\pi\sqrt{2\Lambda_km/N }} \nonumber\\
  & \qquad \inte_0^\infty {_1F_1}\left(l-\frac{K-1}{2},\frac{1}{2},-\frac{uN}{2m}\frac{\bar{r}_k^2}{\Lambda_k}\right)
  \frac{\textrm{d}u}{\sqrt{u\det \left(\mathds{1}_N +uD\right)}} 
\label{sec3.15}
\end{align}
and for $D=\mathds{1}_N$ at
\begin{align}
  \langle p\rangle_{AG}^{(k)} (\bar{r}_k|\Lambda_k,\mathds{1}_N) &= \frac{\Gamma(l-(K-1)/2)\Gamma(l-(K-N)/2)}
                          {\Gamma(l-K/2)\Gamma(N/2)\sqrt{2\pi\Lambda_km/N}}\nonumber\\
  & \qquad\qquad\qquad\qquad 
           U\left(l-\frac{K-1}{2},\frac{1-N}{2}+1,\frac{N}{2m}\frac{\bar{r}_k^2}{\Lambda_k}\right) \ .
\label{sec3.16}
\end{align}
Moreover,
\begin{align}
  \langle p\rangle_{AG}^{(k)} (\bar{r}_k|\Lambda_k,D) &\sim \left(\frac{\bar{r}_k^2}{\Lambda_k}\right)^{-l+(K-1)/2}
\label{sec3.17}
\end{align}
is the asymptotic behavior.

\subsection{Algebraic--Algebraic}
\label{sec35}

Finally, we provide the results for the Algebraic--Algebraic case. In
a non--Markovian situation with $D\neq \mathds{1}_N$, we find after
steps analogous to the ones in Sec.~\ref{sec25}
\begin{align}
  \langle p\rangle_{AA}^{(k)} (\bar{r}_k|\Lambda_k,D) &= \frac{\Gamma(l-(K-1)/2)\Gamma(L-(K+N)/2+1)}
                             {\pi\sqrt{\Lambda_kMm/N}\Gamma(l-K/2)\Gamma(L-(K+N-1)/2)}                  
                  \nonumber\\
  & \ \inte_0^\infty {_2F_1}\left(l-\frac{K-1}{2},L-\frac{K+N}{2}+1,\frac{1}{2},-\frac{vN}{Mm}\frac{\bar{r}_k^2}{\Lambda_k}\right)
                   \frac{\textrm{d}v}{\sqrt{v\det(\mathds{1}_N+vD)}} \ , 
\label{sec3.17a}
\end{align}
while
\begin{align}
  \langle p\rangle_{AA}^{(k)} (\bar{r}_k|\Lambda_k,\mathds{1}_N)
  &= \frac{\Gamma(l-(K-1)/2)\Gamma(l-(K-N)/2)}
                                {\sqrt{\pi\Lambda_kMm/N}\Gamma(l-K/2)\Gamma(L+l-(K-1))} \nonumber\\
  & \ \frac{\Gamma(L-(K-1)/2)\Gamma(L-(K+N)/2+1)}
                                {\Gamma(L-(K+N-1)/2)\Gamma(N/2)} \nonumber\\
     & \  {_2F_1}\left(l-\frac{K-1}{2},L-\frac{K+N}{2}+1,L+l-(K-1),1-\frac{N}{Mm}\frac{\bar{r}_k^2}{\Lambda_k}\right) \ ,
 \label{sec3.18}
\end{align}
results in the Markovian case $D=\mathds{1}_N$. We also obtain
\begin{align}
  \langle p\rangle_{AA}^{(k)} (\bar{r}_k|\Lambda_k,D) &\sim \begin{cases}
                                      \displaystyle
                                      \left(\frac{\bar{r}_k^2}{\Lambda_k}\right)^{-l-(K-1)/2} \ ,     & \text{if } l<L-\frac{N-1}{2}  \\
                                      \displaystyle
                                      \left(\frac{\bar{r}_k^2}{\Lambda_k}\right)^{-l-(K-1)/2} \ln\frac{\bar{r}_k^2}{\Lambda_k} \ ,
                                                                                            & \text{if } l=L-\frac{N-1}{2}  \\
                                      \displaystyle
                                      \left(\frac{\bar{r}_k^2}{\Lambda_k}\right)^{-L+(K+N)/2-1} \ ,     & \text{if } l>L-\frac{N-1}{2} 
                                     \end{cases}
  \label{sec3.19}
\end{align}  
for the asymptotic behavior.

\subsection{Graphical Representations}
\label{sec36}

\begin{figure}[htbp]
  \begin{center}
    \includegraphics[width=0.7\textwidth]{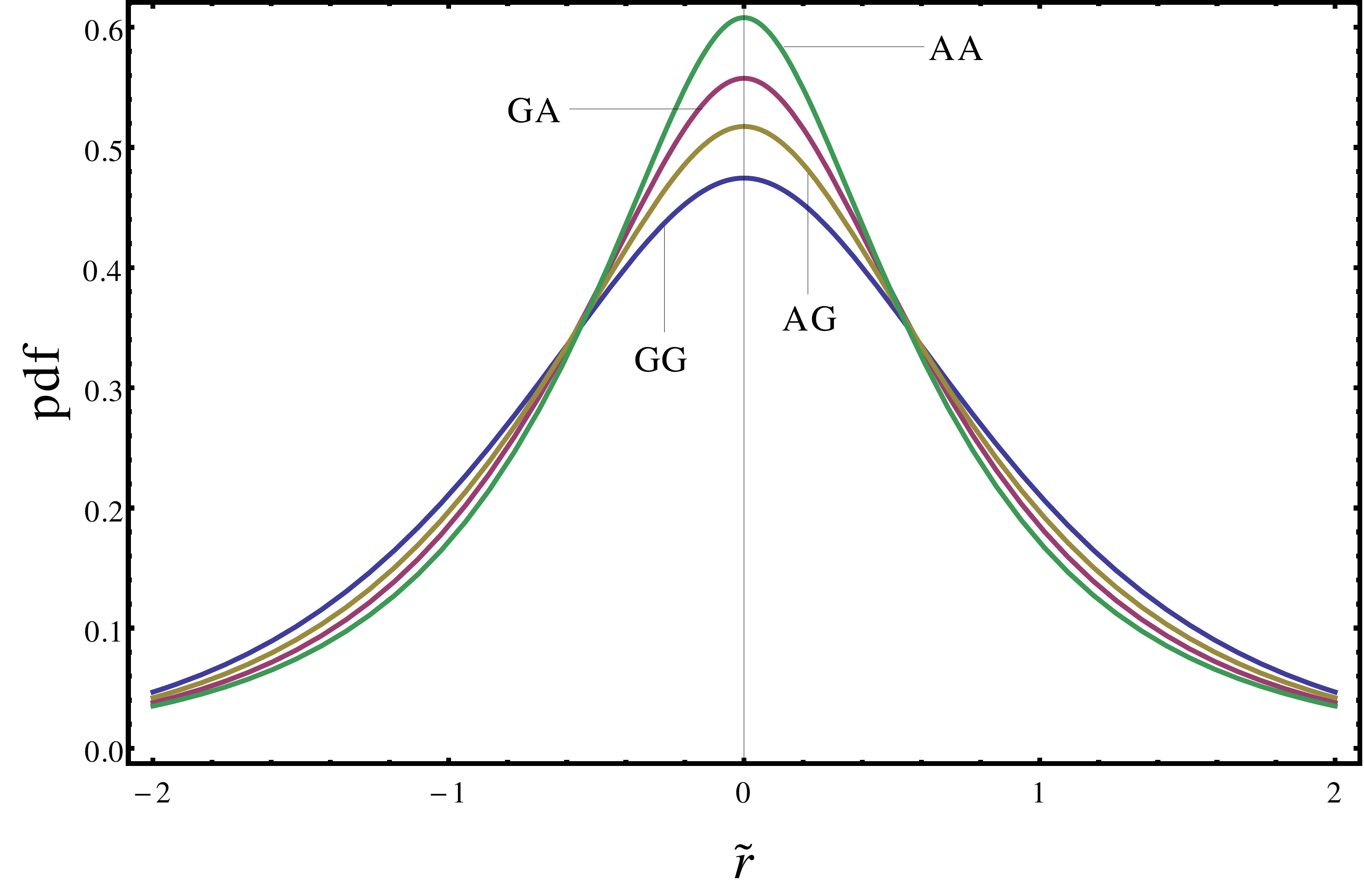}
  \end{center}
  \caption{Probability densities $\langle p\rangle_{YY'}^{(k)}$ in the
    Markovian case versus the rotated amplitudes $\tilde{r}$,
    normalized to unit standard deviation. The four cases
    Gaussian--Gaussian, Gaussian--Algebraic, Algebraic--Gaussian,
    Algebraic--Algebraic are labeled YY' = GG, GA, AG, AA,
    respectively. Number of positions $K=100$, shape parameters $L=55$
    and $l=55$, strength parameter for fluctuations of correlations
    $N=5$. Linear scale, abscissa between -2 and +2.}
 \label{fig3}
\end{figure}

\begin{figure}[htbp]
  \begin{center}
    \includegraphics[width=0.71\textwidth]{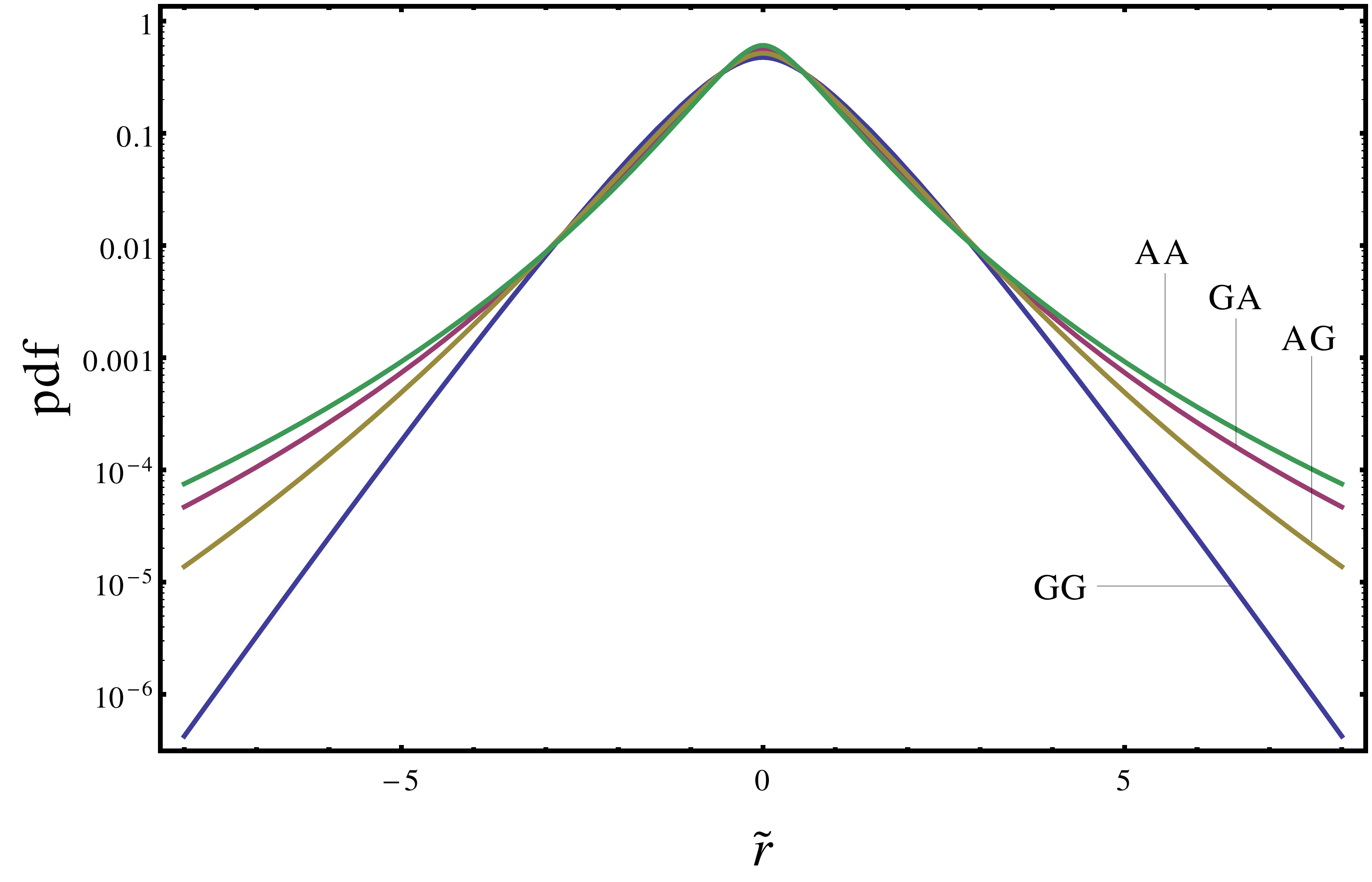}
  \end{center}
  \caption{As Fig.~\ref{fig3}. Logarithmic scale, abscissa between -8 and
    +8.}
 \label{fig4}
\end{figure}

\begin{figure}[htbp]
  \begin{center}
    \includegraphics[width=0.71\textwidth]{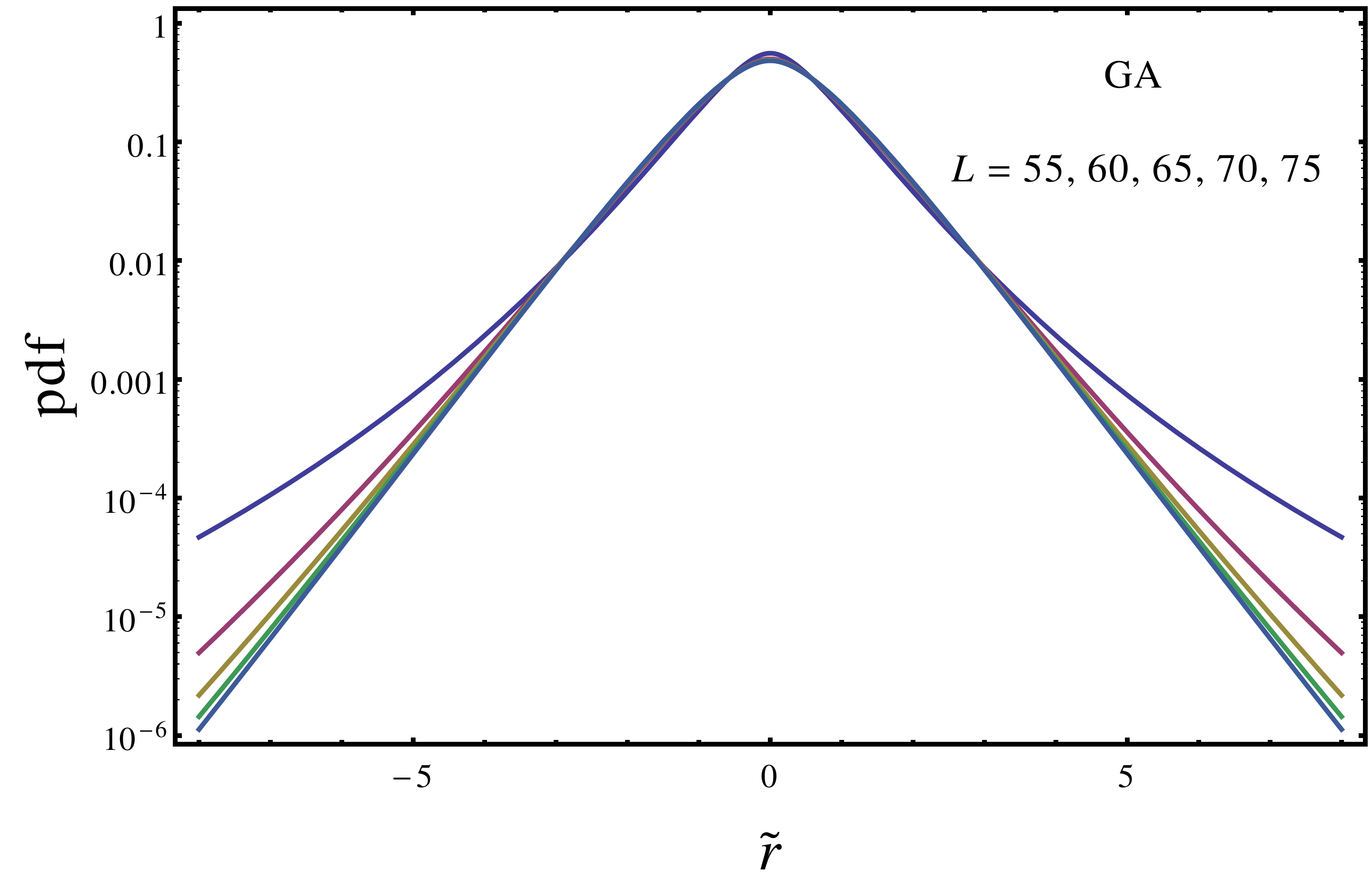}
  \end{center}
  \caption{Probability densities $\langle p\rangle_{GA}^{(k)}$ in the
    Markovian case versus the rotated amplitudes $\tilde{r}$,
    normalized to unit standard deviation, $K=100$, $N=5$, $L$ varying
    as indicated, the smaller $L$, the heavier the tail. Logarithmic
    scale.}
 \label{fig5}
\end{figure}

\begin{figure}[htbp]
  \begin{center}
    \includegraphics[width=0.71\textwidth]{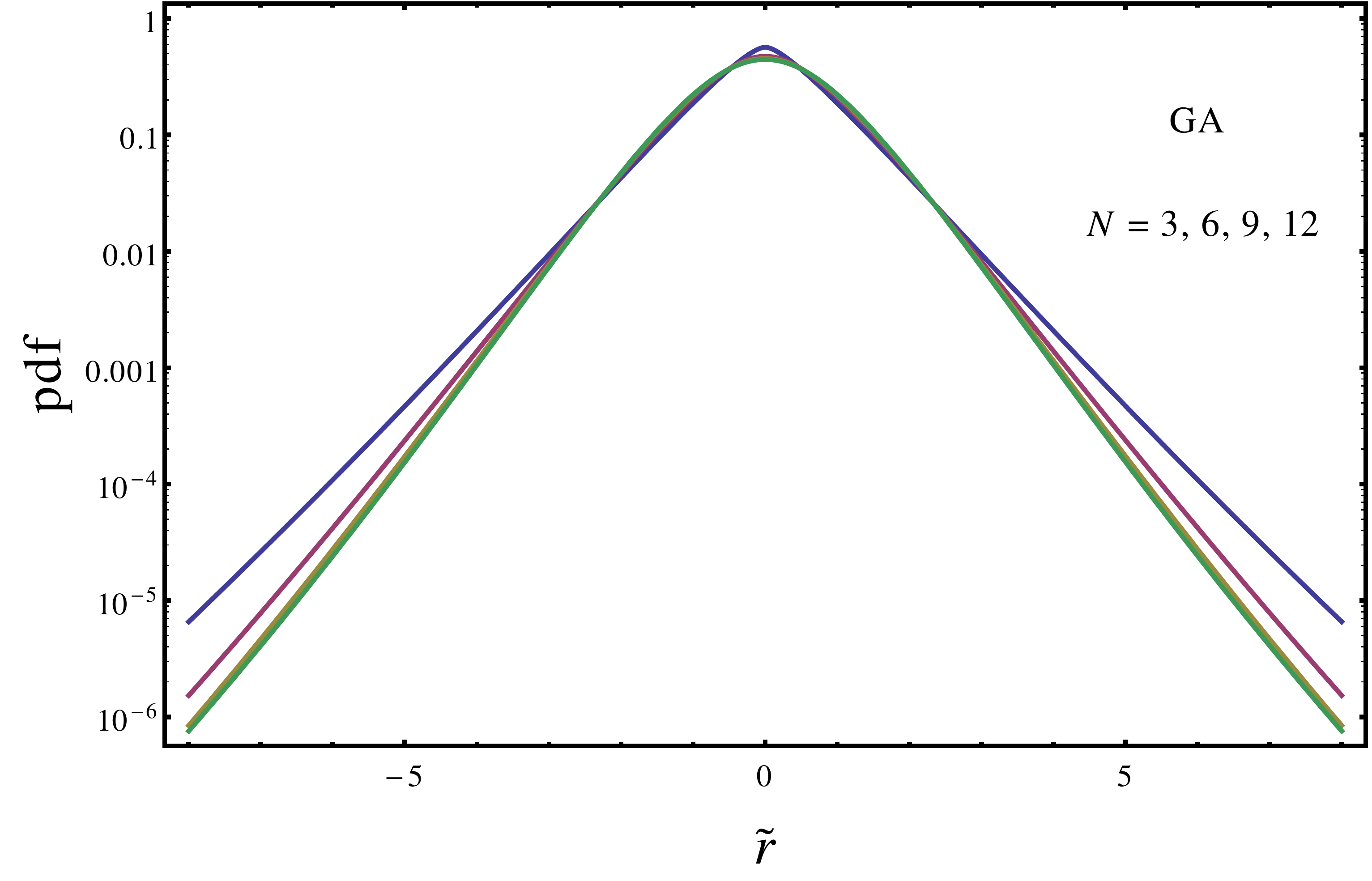}
  \end{center}
  \caption{Probability densities $\langle p\rangle_{GA}^{(k)}$ in the
    Markovian case versus the rotated amplitudes $\tilde{r}$,
    normalized to unit standard deviation, $K=100$, $L=65$, $N$
    varying as indicated, the smaller $N$, the heavier the
    tail. Logarithmic scale.}
 \label{fig6}
\end{figure}

To render possible a comparison of the results involving algebraic
distributions with those in the Gaussian--Gaussian case, we choose
values of $L$ and $l$ which ensure the existence of the first matrix
moment, see Sec.~\ref{sec14}. We notice that the conditions on the
existence of the algebraic distributions, \textit{i.e.}, of their
normalizations, are slightly weaker. We also use the
fixings~\eqref{eq:GauVerM2} and~\eqref{eq:DetVerXm2}, implying that
$\beta_A$ and $B_A$ become one and making sure that the ensemble
averaged correlation matrix, corresponding to the sample correlation
matrix, coincides in all cases with the input matrix $C$. Thus,
according to Eq.~\eqref{sec3.9a} the variances $\langle \bar{r}_k^2
\rangle_{YY'}^{(k)}$ are simply given by $\Lambda_k$. The functional
form of all distributions $\langle p\rangle_{YY'}^{(k)}(\bar{r}_k|D)$
then allows us to normalize the rotated amplitude $\bar{r}_k$ by the
standard deviation,
\begin{align}
  \tilde{r} = \frac{\bar{r}_k}{\sqrt{\Lambda_k}} \ ,
\label{sec3.30}
\end{align}
such that all $K$ distributions in this variable coincide and the
corresponding variances are all given by one. For the graphical
representation, it is useful to look at the Markovian case. We
consider $K=100$ positions.  In Figs.~\ref{fig3} and~\ref{fig4}, we
use the shape parameters $L=55$
and $l=55$, as well as $N=5$ which is a typical value from an empirical
viewpoint.  As seen in Fig.~\ref{fig3}, the more algebraic, the
stronger peaked is the distribution, and as Fig.~\ref{fig4} shows,
the heavier are the tails, corroborating the construction of our
model. In Figs.~\ref{fig5} and~\ref{fig6}, we look at the
Gaussian--Algebraic case for $K=100$ more closely by varying the
dependence on $L$ and $N$. As $L$ is increased for fixed $N=5$, the
convergence of the ensemble distribution for the random data matrix to
the Gaussian competes with the algebraic tail.  Put differently, the
exponential tail known from
the Gaussian--Gaussian case seems to win for larger $L$. But this is
not so, it depends on the range in $\tilde{r}$ considered. On a
sufficiently large $\tilde{r}$ scale, the algebraic tail wins.
Only in the hard limit $L\rightarrow\infty$ for fixed $\tilde{r}$,
there is an exponential tail everywhere. Furthermore, as seen in
Fig.~\ref{fig6}, the smaller the parameter $N$ for fixed $L=65$, the
stronger are the non--stationary fluctuations of the random data
matrices. Of course, this must have a large impact on the
distribution, but it is interesting to observe that the effect
saturates quickly, the curves for $N=9$ and $N=12$ are almost on top
of each other.

\section{Conclusions}
\label{sec4}

We put forward an approach to model multivariate amplitude
distributions in the presence of non--stationarities that imply
fluctuations of correlations. This is a frequently encountered
situation in complex systems. Our model is based on a division of time
scales, the idea behind and the necessary assumption is the existence
of a shorter time scale on which the system is in good approximation
stationary. Longer time scales are then viewed as assembled from
several or many intervals of the shorter time scale.  We argued that
the correlation or covariance matrices measured in all these shorter
time intervals, referred to as epochs, form a truly existing ensemble
that we modeled by an ensemble of random matrices. Averaging the
multivariate amplitude distributions of the epochs over this ensemble
yields heavy tailed amplitude distributions for the larger time scales
on which the non--stationarity is crucial. In short, the fluctations
of the correlations lift the tails of the multivariate amplitude
distributions.

We evaluated our model explicitly for four cases, combining Gaussian
and algebraic multivariate amplitude distributions in the epochs with
Gaussian and algebraic distributions for the random correlation or
covariance matrices. It is a welcome feature that we arrive in all
cases at closed form results if the system is Markovian, but even if
it is not, the resulting formulae are single integrals.  A highly
appreciated effect, as in every ensemble approach, is the reduction in
the number of relevant parameters. In all of our distributions, there
is one parameter that measures the strength with which the
correlations fluctuate. Hence, there is a direct measure for the
non--stationarity that is obtained by fitting the multivariate
distributions to the data. The shape parameters of the algebraic
distributions are best determined by fitting the tails.  For the
distribution for the data in the epochs, the corresponding shape
parameters can be obtained before turning to the model on the large
time scale.

This discussion also explains why we find simple fit formulae, as
often used in finance and referred to as ``models'',
insufficient. Their parameters neither give a clue on underlying
mechanisms nor can they be measured by other means from the data --- unlike
correlations and covariances. From this point of view, our approach
provides justifications of and explanations for fit formulae. Here, we
not only gave explicit multivariate distributions for non--stationary
systems, but furthermore a tool that quantitatively captures the
degree of non--stationarity. Applications to data will be presented in
forthcoming studies.

\section*{Acknowledgement}

We thank Mario Kieburg and Shanshan Wang for helpful discussions.

\appendix
\label{sec:anhang}

\section{Asymptotics of the Ensemble Averaged Amplitude Distributions}
\label{app1}

We always write $w=\sqrt{r^\dagger C^{-1}r}$, we begin with the
Gaussian--Algebraic case of Sec.~\ref{sec23}. To study the integral in
Eq.~\eqref{ga9}, to which we refer as $Q_{GA}$, we write
the determinant as Gaussian integral using an $N$ component vector
$\zeta$ and employ the power series of the Kummer function
\begin{align}
Q_{GA} &= \inte_0^\infty \frac{\text{d}u u^{K/2-1}}{\sqrt{\det(\mathds{1}_N +u D)}} \,
        _1F_1 \left( L-\frac{N-1}{2},\frac{K}{2},-\frac{Nw^2}{2M} u \right)\nonumber\\
&=\frac{1}{\pi^{N/2}} \summ_{i=0}^\infty \frac{\Gamma(L-(N-1)/2 +i)\Gamma(K/2)}
               {\Gamma(L-(N-1)/2)\Gamma(K/2+i)i!} \left(-\frac{Nw^2}{2M} \right)^i\nonumber\\
  &\qquad\qquad\qquad \int\text{d}[\zeta] \exp(-\zeta^\dagger\zeta) \inte_0^\infty \text{d}u
                             u^{K/2+i-1} \exp(-u \zeta^\dagger D\zeta) \ .
\label{app1.1}
\end{align}
The $u$ integral can now be calculated, and we resum the resulting series 
\begin{align}
Q_{GA} &= \frac{\Gamma(K/2)}{\pi^{N/2}} \int\text{d}[\zeta]\exp(-\zeta^\dagger\zeta) (\zeta^\dagger D\zeta)^{-K/2}
             \left(1+\frac{Nw^2}{2M} \frac{1}{\zeta^\dagger D \zeta}\right)^{-L+\frac{N-1}{2}}\nonumber\\
  &= \frac{\Gamma(K/2)}{\sqrt{\det \pi D}} \left( \frac{Nw^2}{2M}\right)^{(N-K)/2}\nonumber\\
  &\qquad\qquad \int \text{d}[\eta] \exp\left(-\frac{Nw^2}{2M} \eta^\dagger D^{-1} \eta\right) (\eta^\dagger \eta)^{L-(N+K-1)/2} (1+\eta^\dagger \eta)^{-L+(N-1)/2} \ ,
\label{app1.2}
\end{align}
where we changed variables according to
$\zeta=\sqrt{Nw^2/2M}D^{-1/2}\eta$ in the second equation.
Introducing hyperspherical coordinates $\eta = \rho e_\eta$ with
$\text{d}[\eta]=\rho^{N-1}\textrm{d}\rho\textrm{d}\Omega_N$ where
$\textrm{d}\Omega_N$ is the infinitesimal solid angle in $N$
dimensions, we identify the $\rho$ integral as the integral
representation~\eqref{bt5} of the Tricomi function
\begin{align}
  Q_{GA} &=   \frac{\Gamma(K/2)\Gamma(L-(K-1)/2)}{2\sqrt{\det \pi D}}
        \left( \frac{N w^2}{2M} \right)^{(N-K)/2}\nonumber\\
  &\qquad\qquad\qquad \int U\left(L-\frac{K-1}{2},\frac{N-K}{2}
        +1,\frac{Nw^2}{2M} e_\eta^\dagger D^{-1}e_\eta\right)\textrm{d}\Omega_N \ .
\label{app1.3}
\end{align}
The Tricomi function has an asymptotic expansion in its last argument
which yields as leading term the inverse of this last argument raised
to the power of its first argument. Thus we find
\begin{align}
  \langle p\rangle_{GA} (r|C,D) &\longrightarrow 
  \frac{\Gamma(L-(N-1)/2)\Gamma(L-(K-1)/2)}{2\Gamma(L-(K+N-1)/2)\sqrt{\det2\pi CM/N }\sqrt{\det \pi D}}\nonumber\\
    &\qquad\qquad \left( \frac{Nw^2}{2M}\right)^{-L+\frac{N-1}{2}}
  \int (e_\eta^\dagger D^{-1} e_\eta)^{-L+(K-1)/2}\textrm{d}\Omega_N \ .
\label{app1.4}  
\end{align}
Most conveniently, the angular average appears only as a factor
independent of $w$ and we arrive, as claimed, at the asymptotic
behavior~\eqref{ga10} of the ensemble averaged amplitude distribution
for large $w=\sqrt{r^\dagger C^{-1}r}$. Indirectly, this derivation
shows that the ensemble avearged amplitude
distribution~\eqref{eq:GauVerR2} in the case of Gaussian distributed
random data matrices cannot be algebraic.  As discussed in
Sec.~\ref{sec23}, the limit $L,M\to\infty$ of $\langle p\rangle_{GA}
(r|C,D)$ under the condition~\eqref{r6} yields $\langle p\rangle_{GG}
(r|C,D)$. This limit, however, is in conflict with the above used
asymptotic expansion of the Tricomi function. If such an expansion
cannot be applied, the angular average will never separate off as a
simple product, seriously hampering a direct derivation of the
exponential asymptotics in the Gaussian--Gaussian case for $D \neq
\mathds{1}_N$.

The algebraic tail in the Algebraic--Gaussian case of Sec.~\ref{sec24}
is derived accordingly. We notice that in this case, too, the tail in
the Markovian situation $D=\mathds{1}_N$ immediately follows from the
results~\eqref{ga10} and~\eqref{ag6} due to the above mentioned
asymptotic expansion of the Tricomi function which features its last
argument to the power of its first argument. With this in mind, we now
turn to the Algebraic--Algebraic case of Sec.~\ref{sec25}. We only
consider the Markovian situation $D=\mathds{1}_N$ as we may expect
that it yields the same asymptotics for the non--Markovian one,
too. Thus we consider in the result~\eqref{aa7} the hypergeometric
function to which we refer as $Q_{AA}$. In the regime $w\gg Mm/N$, we
may use the approximation
\begin{align}
Q_{AA} & \simeq {_2F_1}\left(l,L-\frac{N-1}{2},L+l-\frac{K-1}{2},-\frac{Nr^\dagger C^{-1}r}{Mm}\right) \ .
\label{app1.5}  
\end{align}
An asymptotic expansion can then be infered from the transformation
formula~\cite{Gradshteyn2007}
\begin{align}
&  \frac{\sin\pi(b-a)}{\pi}\frac{{_2F_1}(a,b,c,z)}{\Gamma(c)} =
        \frac{(-z)^{-a}}{\Gamma(b)\Gamma(c-a)\Gamma(a-b+1)}{_2F_1}\left(a,a-c+1,a-b+1,\frac{1}{z}\right) \nonumber\\
& \qquad\qquad\qquad\qquad -\frac{(-z)^{-b}}{\Gamma(a)\Gamma(c-b)\Gamma(b-a+1)}{_2F_1}\left(b,b-c+1,b-a+1,\frac{1}{z}\right) \ .
\label{app1.6}  
\end{align}
The first term in the defining Gauss series of the hypergeometric
function is one, implying that the two hypergeometric functions on the
right hand side become one for large $z$. The asymptotics is hence
determined by $(-z)^{-a}$ or $(-z)^{-b}$, depending on whether $a<b$
or $b>a$. The sine function on the left hand side produces a sign
which ensures the positivity of the distribution. To apply this
particular line of reasoning to $Q_{AA}$, we must assume that $N$ is
even, for odd $N$ cancelations have to prevent the functions
$\Gamma(a-b+1)$ and $\Gamma(a-b+1)$ from diverging. If $b=a$, we bring
the sine function on the right hand side of Eq.~\eqref{app1.6} and use
l'Hospital's rule which yields
\begin{align}
  {_2F_1}(a,b,c,z) &\sim \frac{(-z)^{-a}}{\Gamma(a)\Gamma(c)}
  \left(1+\frac{\ln(-z)-2\psi(1)-\psi(a)-\psi(c-a)}{\Gamma(c-a)z^0}+
  \ldots\right) \ ,
\label{app1.7}  
\end{align}
resulting in a logarithmic correction to the asymptotic behavior.
Here $\psi(z)=\Gamma'(z)/\Gamma(z)$ is the digamma function.


%
%
%

\begin{thebibliography}{10}

\bibitem{Gao1999}
J.~B. Gao.
\newblock {Recurrence Time Statistics for Chaotic Systems and Their
  Applications}.
\newblock {\em Physical Review Letters}, 83:3178--3181, Oct 1999.

\bibitem{Hegger2000}
Rainer Hegger, Holger Kantz, Lorenzo Matassini, and Thomas Schreiber.
\newblock {Coping with Nonstationarity by Overembedding}.
\newblock {\em Physical Review Letters}, 84:4092--4095, May 2000.

\bibitem{Bernaola2001}
Pedro Bernaola-Galv\'an, Plamen~Ch. Ivanov, Lu\'{\i}s~A. Nunes~Amaral, and
  H.~Eugene Stanley.
\newblock {Scale Invariance in the Nonstationarity of Human Heart Rate}.
\newblock {\em Physical Review Letters}, 87:168105, Oct 2001.

\bibitem{Rieke2002}
Christoph Rieke, Karsten Sternickel, Ralph~G. Andrzejak, Christian~E. Elger,
  Peter David, and Klaus Lehnertz.
\newblock {Measuring Nonstationarity by Analyzing the Loss of Recurrence in
  Dynamical Systems}.
\newblock {\em Physical Review Letters}, 88:244102, May 2002.

\bibitem{Zia2004}
R.~Zia and Per~Arne Rikvold.
\newblock {Fluctuations and correlations in an individual-based model of
  biological coevolution}.
\newblock {\em Journal of Physics A}, 37, 02 2004.

\bibitem{Zia2006}
R~K~P Zia and B~Schmittmann.
\newblock {A possible classification of nonequilibrium steady states}.
\newblock {\em Journal of Physics A}, 39(24):L407–L413, may 2006.

\bibitem{pijn1991chaos}
Jan~Pieter Pijn, Jan Van~Neerven, Andr{\'e} Noest, and Fernando H~Lopes
  da~Silva.
\newblock {Chaos or noise in EEG signals; dependence on state and brain site}.
\newblock {\em Electroencephalography and Clinical Neurophysiology},
  79(5):371--381, 1991.

\bibitem{Mueller2005}
Markus M\"uller, Gerold Baier, Andreas Galka, Ulrich Stephani, and Hiltrud
  Muhle.
\newblock {Detection and characterization of changes of the correlation
  structure in multivariate time series}.
\newblock {\em Physical Review E}, 71:046116, Apr 2005.

\bibitem{Hohmann2010}
R.~H\"ohmann, U.~Kuhl, H.-J. St\"ockmann, L.~Kaplan, and E.~J. Heller.
\newblock {Freak Waves in the Linear Regime: A Microwave Study}.
\newblock {\em Physical Review Letters}, 104:093901, Mar 2010.

\bibitem{metzger2014statistics}
Jakob~J. Metzger, Ragnar Fleischmann, and Theo Geisel.
\newblock {Statistics of Extreme Waves in Random Media}.
\newblock {\em Physical Review Letters}, 112:203903, May 2014.

\bibitem{degueldre2015random}
Henri Degueldre, Jakob Metzger, Theo Geisel, and Ragnar Fleischmann.
\newblock {Random Focusing of Tsunami Waves}.
\newblock {\em Nature Physics}, 12, 11 2015.

\bibitem{bekaert1995}
Geert Bekaert and Campbell~R. Harvey.
\newblock {Time-Varying World Market Integration}.
\newblock {\em Journal of Finance}, 50(2):403--444, 1995.

\bibitem{Longin1995}
Francois Longin and Bruno Solnik.
\newblock {Is the correlation in international equity returns constant:
  1960-1990?}
\newblock {\em Journal of International Money and Finance}, 14(1):3--26, 1995.

\bibitem{Onnela2003}
J.-P. Onnela, A.~Chakraborti, K.~Kaski, J.~Kert\'esz, and A.~Kanto.
\newblock {Dynamics of market correlations: Taxonomy and portfolio analysis}.
\newblock {\em Physical Review E}, 68:056110, Nov 2003.

\bibitem{Zhang2011}
Yiting Zhang, Gladys Hui~Ting Lee, Jian~Cheng Wong, Jun~Liang Kok, Manamohan
  Prusty, and Siew~Ann Cheong.
\newblock {Will the US economy recover in 2010? A minimal spanning tree study}.
\newblock {\em Physica A}, 390(11):2020--2050, 2011.

\bibitem{Song2011}
Dong-Ming Song, Michele Tumminello, Wei-Xing Zhou, and Rosario~N. Mantegna.
\newblock {Evolution of worldwide stock markets, correlation structure, and
  correlation-based graphs}.
\newblock {\em Physical Review E}, 84:026108, Aug 2011.

\bibitem{Sandoval2012}
Leonidas Sandoval and Italo De~Paula Franca.
\newblock {Correlation of financial markets in times of crisis}.
\newblock {\em Physica A}, 391(1):187--208, 2012.

\bibitem{Munnix2012}
Michael~C. M\"unnix, Takashi Shimada, Rudi Sch\"afer, Francois Leyvraz,
  Thomas~H. Seligman, Thomas Guhr, and H.~Eugene Stanley.
\newblock {Identifying States of a Financial Market}.
\newblock {\em Scientific Reports}, 2:644, Sep 2012.

\bibitem{Ghasemi2006}
F.~Ghasemi, Muhammad Sahimi, Joachim Peinke, and M~Tabar.
\newblock {Analysis of Non-stationary Data for Heart-Rate Fluctuations in Terms
  of Drift and Diffusion Coefficients}.
\newblock {\em Journal of Biological Physics}, 32:117--28, 11 2006.

\bibitem{PhysRevE.87.062139}
Mehrnaz Anvari, Cina Aghamohammadi, H.~Dashti-Naserabadi, E.~Salehi, E.~Behjat,
  M.~Qorbani, M.~{Khazaei Nezhad}, M.~Zirak, Ali Hadjihosseini, Joachim Peinke,
  and M.~Reza~Rahimi Tabar.
\newblock {Stochastic nature of series of waiting times}.
\newblock {\em Physical Review E}, 87:062139, Jun 2013.

\bibitem{PhysRevE.75.060102}
Fatemeh Ghasemi, Muhammad Sahimi, J.~Peinke, R.~Friedrich, G.~Reza Jafari, and
  M.~Reza~Rahimi Tabar.
\newblock {Markov analysis and Kramers-Moyal expansion of nonstationary
  stochastic processes with application to the fluctuations in the oil price}.
\newblock {\em Physical Review E}, 75:060102, Jun 2007.

\bibitem{Schafer20103856}
Rudi Sch{\"a}fer and Thomas Guhr.
\newblock {Local normalization: Uncovering correlations in non-stationary
  financial time series}.
\newblock {\em Physica A}, 389(18):3856--3865, 2010.

\bibitem{bohr1969nuclear}
A.~Bohr and B.R. Mottelson.
\newblock {\em {Nuclear Structure: Volume 2, Nuclear deformations}}.
\newblock Nuclear Structure. Benjamin, 1969.

\bibitem{zelevinsky1996quantum}
Vladimir Zelevinsky.
\newblock {Quantum Chaos and Complexity in Nuclei}.
\newblock {\em Annual Review of Nuclear and Particle Science}, 46(1):237--279,
  1996.

\bibitem{Guhr1998}
Thomas Guhr, Axel M\"uller-Groeling, and Hans~A. Weidenm\"uller.
\newblock {Random Matrix Theories in Quantum Physics: Common Concepts}.
\newblock {\em Physics Reports}, 299:189--425, 1998.

\bibitem{Schmitt2013}
Thilo~A. Schmitt, Desislava Chetalova, Rudi Sch{\"a}fer, and Thomas Guhr.
\newblock {Non-stationarity in financial time series: Generic features and tail
  behavior}.
\newblock {\em Europhysics Letters}, 103(5):58003, 2013.

\bibitem{Meudt2015}
Frederik Meudt, Martin Theissen, Rudi Sch\"afer, and Thomas Guhr.
\newblock {Constructing Analytically Tractable Ensembles of Non-Stationary
  Covariances with an Application to Financial Data}.
\newblock {\em Journal of Statistical Mechanics}, {2015(11)}:P11025, 03 2015.

\bibitem{schmitt2014credit}
Thilo Schmitt, Desislava Chetalova, Rudi Sch\"afer, and Thomas Guhr.
\newblock {Credit Risk and the Instability of the Financial System: an Ensemble
  Approach}.
\newblock {\em Europhysics Letters}, 105, 09 2013.

\bibitem{schmitt2015credit}
Thilo Schmitt, Rudi Sch\"afer, and Thomas Guhr.
\newblock {Credit Risk: Taking Fluctuating Asset Correlations into Account}.
\newblock {\em Journal of Credit Risk}, 11:73--94, 09 2015.

\bibitem{SBGSK2015}
Rudi Sch\"afer, Sonja Barkhofen, Thomas Guhr, Hans-J\"urgen St\"ockmann, and
  Ulrich Kuhl.
\newblock {Compounding Approach for Univariate Time Series with Nonstationary
  Variances}.
\newblock {\em Physical Review E}, 92:062901, Dec 2015.

\bibitem{dubey1970compound}
Satya Dubey.
\newblock {Compound gamma, beta and F distributions}.
\newblock {\em Metrika: International Journal for Theoretical and Applied
  Statistics}, 16(1):27--31, 1970.

\bibitem{10.2307/1402598}
O.~Barndorff-Nielsen, J.~Kent, and M.~Sørensen.
\newblock {Normal Variance-Mean Mixtures and z Distributions}.
\newblock {\em International Statistical Review / Revue Internationale de
  Statistique}, 50(2):145--159, 1982.

\bibitem{Beck2003}
C.~Beck and E.G.D. Cohen.
\newblock {Superstatistics}.
\newblock {\em Physica A}, 322(C):267--275, 2003.

\bibitem{Abul-Magd2009}
A.~Y. Abul-Magd, G.~Akemann, and P.~Vivo.
\newblock {Superstatistical generalizations of Wishart--Laguerre ensembles of
  random matrices}.
\newblock {\em Journal of Physics A}, 42(17):175207, 2009.

\bibitem{Doulgeris2010}
Anthony~Paul Doulgeris and Torbj{\o}rn Eltoft.
\newblock {Scale Mixture of Gaussian Modelling of Polarimetric {SAR} Data}.
\newblock {\em {EURASIP} Journal on Advances in Signal Processing}, 2010, 2010.

\bibitem{Forbes2014}
Florence Forbes and Darren Wraith.
\newblock {A new family of multivariate heavy-tailed distributions with
  variable marginal amounts of tailweight: Application to robust clustering}.
\newblock {\em Statistics and Computing}, 24, 11 2014.

\bibitem{bouchaud2000theory}
JP~Bouchaud and M~Potters.
\newblock {\em {Theory of Financial Risks, From Statistical Physics to Risk
  Management}}.
\newblock Cambridge University Press, New York, 2000.

\bibitem{Mehta2004}
M.L. Mehta.
\newblock {\em {Random Matrices}}.
\newblock ISSN. Elsevier Science, 2004.

\bibitem{Wishart1928}
J.~Wishart.
\newblock {The Generalised Product Moment Distribution in Samples From a Normal
  Multivariate Population}.
\newblock {\em Biometrika}, 20A(1-2):32--52, 1928.

\bibitem{Laloux1999}
Laurent Laloux, Pierre Cizeau, Jean-Philippe Bouchaud, and Marc Potters.
\newblock {Noise Dressing of Financial Correlation Matrices}.
\newblock {\em Physical Review Letters}, 83:1467--1470, Aug 1999.

\bibitem{Laloux2000}
Laurent Laloux, Pierre Cizeau, Marc Potters, and Jean-Philippe Bouchaud.
\newblock {Random Matrix Theory and Financial Correlations}.
\newblock {\em International Journal of Theoretical and Applied Finance},
  03(03):391--397, 2000.

\bibitem{Plerou1999a}
Vasiliki Plerou, Parameswaran Gopikrishnan, Bernd Rosenow, Lu\'{\i}s~A.
  Nunes~Amaral, and H.~Eugene Stanley.
\newblock {Universal and Nonuniversal Properties of Cross Correlations in
  Financial Time Series}.
\newblock {\em Physical Review Letters}, 83:1471--1474, Aug 1999.

\bibitem{Plerou2002}
Vasiliki Plerou, Parameswaran Gopikrishnan, Bernd Rosenow, Luis A.~Nunes
  Amaral, Thomas Guhr, and H.~Eugene Stanley.
\newblock {Random Matrix Approach to Cross Correlations in Financial Data}.
\newblock {\em Physical Review E}, 65:066126, Jun 2002.

\bibitem{Pafka2004}
Szil\'ard Pafka and Imre Kondor.
\newblock {Estimated correlation matrices and portfolio optimization}.
\newblock {\em Physica A}, 343(C):623--634, 2004.

\bibitem{potters2005financial}
Marc Potters, Jean-Philippe Bouchaud, and Laurent Laloux.
\newblock {Financial Applications of Random Matrix Theory: Old Laces and New
  Pieces}.
\newblock {\em Acta Physica Polonica B}, 36:2767--2784, 09 2005.

\bibitem{drozdz2008empirics}
S~Drozdz, J~Kwapien, and P~Oswiecimka.
\newblock {Empirics Versus RMT in Financial Cross-Correlations}.
\newblock {\em Acta Physica Polonica B}, 39(1):4027--4039, 2008.

\bibitem{Kwapien2006}
Jaroslaw Kwapien, Stanislaw Drozdz, and P.~Oswiecimka.
\newblock {The bulk of the stock market correlation matrix is not pure noise}.
\newblock {\em Physica A}, 359:589--606, 05 2006.

\bibitem{biroli2007student}
Giulio Biroli, Jean-Philippe Bouchaud, and Marc Potters.
\newblock {The Student ensemble of correlation matrices: eigenvalue spectrum
  and Kullback-Leibler entropy}.
\newblock {\em Acta Physica Polonica B}, 38:4009--4026, 10 2007.

\bibitem{Burda2001}
Zdzislaw Burda, J.~Jurkiewicz, Maciej Nowak, G\'abor Papp, and Ismail Zahed.
\newblock {Free Levy Matrices and Financial Correlations}.
\newblock {\em Physica A}, 343:694--700, 04 2001.

\bibitem{Burda2002}
Zdzis\l{}aw Burda, Romuald~A. Janik, Jerzy Jurkiewicz, Maciej~A. Nowak, Gabor
  Papp, and Ismail Zahed.
\newblock {Free random L\'evy matrices}.
\newblock {\em Physical Review E}, 65:021106, Jan 2002.

\bibitem{Akemann2008}
Gernot Akemann and Pierpaolo Vivo.
\newblock {Power law deformation of Wishart--Laguerre ensembles of random
  matrices}.
\newblock {\em Journal of Statistical Mechanics}, 2008(09):P09002, 2008.

\bibitem{burda2011applying}
Zdzislaw Burda, Andrzej Jarosz, {Maciej A.} Nowak, Jerzy Jurkiewicz, G{\'a}bor
  Papp, and Ismail Zahed.
\newblock {Applying free random variables to random matrix analysis of
  financial data. Part I: The Gaussian case}.
\newblock {\em Quantitative Finance}, 11(7):1103--1124, Jul 2011.

\bibitem{French1978}
J.B. French, P.A. Mello, and A.~Pandey.
\newblock {Ergodic behavior in the statistical theory of nuclear reactions}.
\newblock {\em Physics Letters B}, 80:17--19, 1978.

\bibitem{Haq1982}
R.~U. Haq, A.~Pandey, and O.~Bohigas.
\newblock {Fluctuation Properties of Nuclear Energy Levels: Do Theory and
  Experiment Agree?}
\newblock {\em Physical Review Letters}, 48:1086--1089, Apr 1982.

\bibitem{Jac00}
J.J.M. Verbaarschot and T.~Wettig.
\newblock {Random Matrix Theory and Chiral Symmetry in QCD}.
\newblock {\em Annual Review of Nuclear and Particle Science}, 50(1):343--410,
  2000.

\bibitem{Mahalanobis36}
P.C. Mahalanobis.
\newblock {On the generalised distance in statistics}.
\newblock {\em Proceedings of the National Institute of Science of India},
  2:49--55, 1936.

\bibitem{Simon2004}
Steven~H. Simon and Aris~L. Moustakas.
\newblock {Eigenvalue density of correlated complex random Wishart matrices}.
\newblock {\em Physical Review E}, 69:065101, Jun 2004.

\bibitem{Burda2005}
Zdzis\l{}aw Burda, Jerzy Jurkiewicz, and Bart\l{}omiej Wac\l{}aw.
\newblock {Spectral moments of correlated Wishart matrices}.
\newblock {\em Physical Review E}, 71:026111, Feb 2005.

\bibitem{McKay2007}
M.~R. {McKay}, A.~J. {Grant}, and I.~B. {Collings}.
\newblock {Performance Analysis of MIMO-MRC in Double-Correlated Rayleigh
  Environments}.
\newblock {\em IEEE Transactions on Communications}, 55(3):497--507, 2007.

\bibitem{Waltner2014}
Daniel Waltner, Tim Wirtz, and Thomas Guhr.
\newblock {Eigenvalue Density of the Doubly Correlated Wishart Model: Exact
  Results}.
\newblock {\em Journal of Physics A}, 48:175204, 12 2014.

\bibitem{Burda2015}
Zdzislaw Burda and Jerzy Jurkiewicz.
\newblock {Chapter 13, Heavy-tailed Random Matrices}.
\newblock In Gernot Akemann, Jinho Baik, and Philippe Di~Francesco, editors,
  {\em Oxford Handbook of Random Matrix Theory}, page 270. Oxford University
  Press, 09 2015.

\bibitem{FK2009}
Peter Forrester and Manjunath Krishnapur.
\newblock {Derivation of an eigenvalue probability density function relating to
  the Poincare disk}.
\newblock {\em Journal of Physics A}, 42, 2009.

\bibitem{WWKK2016}
Tim Wirtz, Daniel Waltner, Mario Kieburg, and Santosh Kumar.
\newblock {The Correlated Jacobi and the Correlated Cauchy-Lorentz ensembles}.
\newblock {\em Journal of Statistical Physics}, 162:495, 01 2016.

\bibitem{GS2020b}
Thomas Guhr and Andreas Schell.
\newblock {Matrix Moments in a Real, Doubly Correlated Algebraic Generalization
  of the Wishart Model}.
\newblock {\em submitted for publication}, 2020.

\bibitem{Siegel_1935}
Carl~Ludwig Siegel.
\newblock {\"Uber Die Analytische Theorie Der Quadratischen Formen}.
\newblock {\em Annals of Mathematics}, 36(3):527--606, 1935.

\bibitem{Fyodorov_2002}
Yan~V. Fyodorov.
\newblock {Negative moments of characteristic polynomials of random matrices:
  Ingham--Siegel integral as an alternative to Hubbard--Stratonovich
  transformation}.
\newblock {\em Nuclear Physics B}, 621(3):643--674, Jan 2002.

\bibitem{Gradshteyn2007}
I.S. Gradsteyn and I.M.Ryzhik.
\newblock {\em {Table of Integrals, Series, and Products}}.
\newblock Academic Press, 7 edition, 2007.

\end{thebibliography}
%


\end{document}